\documentclass[floatfix,reprint,prl,aps,superscriptaddress]{revtex4-2}
\usepackage{times}
\usepackage{graphicx}
\usepackage{amsmath, braket, amsfonts}
\usepackage{amssymb}
\usepackage{bm, color, ulem}
\usepackage{xcolor}
\usepackage{ragged2e}
\usepackage{wrapfig,lipsum,booktabs}
\usepackage{lineno}

\newcommand{\UCSB}{Department of Physics, University of California at Santa Barbara, Santa Barbara, California 93106, USA}
\newcommand{\nimsTT}{Research Center for Materials Nanoarchitectonics, National Institute for Materials Science, 1-1 Namiki, Tsukuba 305-0044, Japan}
\newcommand{\nimsKW}{
Research Center for Electronic and Optical Materials, National Institute for Materials Science, 1-1 Namiki, Tsukuba 305-0044, Japan}
\newcommand{\Berkeley}{Department of Physics, University of California, Berkeley, California 94720, USA}
\newcommand{\BerkeleyMaterial}{Material Science Division, Lawrence Berkeley National Laboratory, Berkeley, California 94720, USA}
\newcommand{\MIT}{Department of Physics, Massachusetts Institute of Technology, Cambridge, MA-02139, USA}
\newcommand{\Harvard}{Department of Physics, Harvard University, Cambridge, MA 02138, USA}
 
\usepackage{siunitx}
\newcommand{\abs}[1]{\left\lvert #1 \right\rvert}

\newcommand{\be}{\begin{equation}}
\newcommand{\ee}{\end{equation}}

\usepackage{mathtools}
\usepackage{physics}

\DeclareUnicodeCharacter{03BD}{{$\nu$}}
\AtBeginDocument{\RenewCommandCopy\qty\SI}
\AtBeginDocument{\RenewCommandCopy\qty\SI} 
\makeatletter
\def\p@figure{Fig.~}
\def\p@table{Table~}
\makeatother
\begin{document}

\title{Single anyons and anyon binding in a quantum Hall dot}
\author{Marco Valentini}
\email{These authors contributed equally}
\author{Will Wang}
\email{These authors contributed equally}
\affiliation{\UCSB}
\author{Amir Shapour Mohammadi}
\email{These authors contributed equally}
\affiliation{\Berkeley}
\author{Noah Samuelson}
\author{Chengxuan Li}
\author{Liam Cohen}
\author{Mark Zakharyan}
\affiliation{\UCSB}
\author{Jiechao Feng}
\affiliation{\Berkeley}
\author{Taige Wang}
\affiliation{\MIT}
\affiliation{\Harvard}
\author{Takashi Taniguchi}
\affiliation{\nimsTT}
\author{Kenji Watanabe}
\affiliation{\nimsKW}
\author{Michael P. Zaletel}
\affiliation{\Berkeley, \BerkeleyMaterial}
\author{Andrea F. Young}
\email{andrea@physics.ucsb.edu}
\affiliation{\UCSB}
\date{\today}
\maketitle 
\onecolumngrid
\begin{center}
\begin{minipage}{0.92\textwidth}
\end{minipage}%
\end{center}
\twocolumngrid

\textbf{
Anyons are particles with quantized fractional statistics.  
In theory, a single anyon can be stabilized as a localized fractional charge within the bulk of a topologically ordered two-dimensional insulator such as a fractional quantum Hall phase.  
However, experiments to date have focused on probing anyons in structures in which many anyonic excitations coexist and interact. 
Here, we demonstrate the trapping of a single anyon in a device consisting of a graphene-based quantum point contact with an embedded, electric-field-effect tunable quantum Hall dot. 
Benchmarking our device at $\nu=1$, we demonstrate backscattering between Fermi liquid edge states mediated by the whole-electron Coulomb blockade of the dot. 
The cleanliness and control available in our device allow depletion of the dot to the last quasihole, and we observe robust spin blockade for transitions to the opposite-spin Landau level.  
At $\nu=1/3$ and $2/3$, we observe backscattering resonances with three times the frequency, consistent with resonant transmission of $e/3$ charges via a `single anyon transistor.' We demonstrate controlled depletion to the last quasiparticle to form a single anyon quantum dot.  Tracing the charge addition spectrum for quasiparticle number $N_{\rm qp}=\text{1-5}$, we observe highly irregular addition energies, consistent with density matrix renormalization group modeling showing a complex orbital shell-filling behavior within the anyonic dot. 
At higher $|N_{\rm qp}|\approx \text{10-20}$, the addition energies show a striking pattern: quasiholes at $1/3$ (and quasiparticles at $2/3$) show a regular addition spectrum while quasiparticles at $1/3$ (and quasiholes at $2/3$) do not. 
We trace the origin of this irregularity to an attractive component of the inter-anyon interaction:  by experimentally tuning the confining energy, we observe complete suppression of alternating charge sectors, after which the dot charges in increments of twice the fundamental charge. 
This observation indicates the emergence of an anyon binding energy at least as large as the charging energy.}

Tremendous progress has been made in elucidating the basic properties of anyons in fractional quantum Hall (FQH) systems\cite{feldman_fractional_2013} since the theoretical underpinnings of the field were established in the 1980s\cite{laughlin_anomalous_1983,halperin_statistics_1984}.  
These experiments have focused on many-anyon ensembles in mesoscopic devices including both large dots\cite{goldman_resonant_1995} and interferometers\cite{ji_electronic_2003,mcclure_fabry-perot_2012}. 
At $\nu=1/3$, for example, the fractional charge has been established by measurements of the addition energy in micron-scale confined geometries arising either from sample disorder\cite{martin_localization_2004} or gate-defined potentials\cite{roosli_fractional_2021}, as well as by shot noise \cite{de_picciotto_direct_1997,saminadayar_observation_1997}. 
The fractional phase for a localized anyon was established  by measuring the relative change in phase for anyon addition in Fabry-Perot\cite{nakamura_direct_2020, werkmeister_anyon_2025, samuelson_slow_2026} and Mach-Zehnder\cite{ghosh_anyonic_2025} interferometer geometries, again in the many bulk quasiparticle regime. A separate line of attack has leveraged scanning tunneling microscopy and the exposed surface of a graphene electron gas to probe localized states in a FQH bulk\cite{deng_real-space_2026,park_local_2026}, although these experiments are limited by the need to tunnel whole electrons into the fractional Hall system, preventing direct spectroscopy of a single anyon localized state.

\begin{figure*}[t]
    \centering
    \includegraphics[]{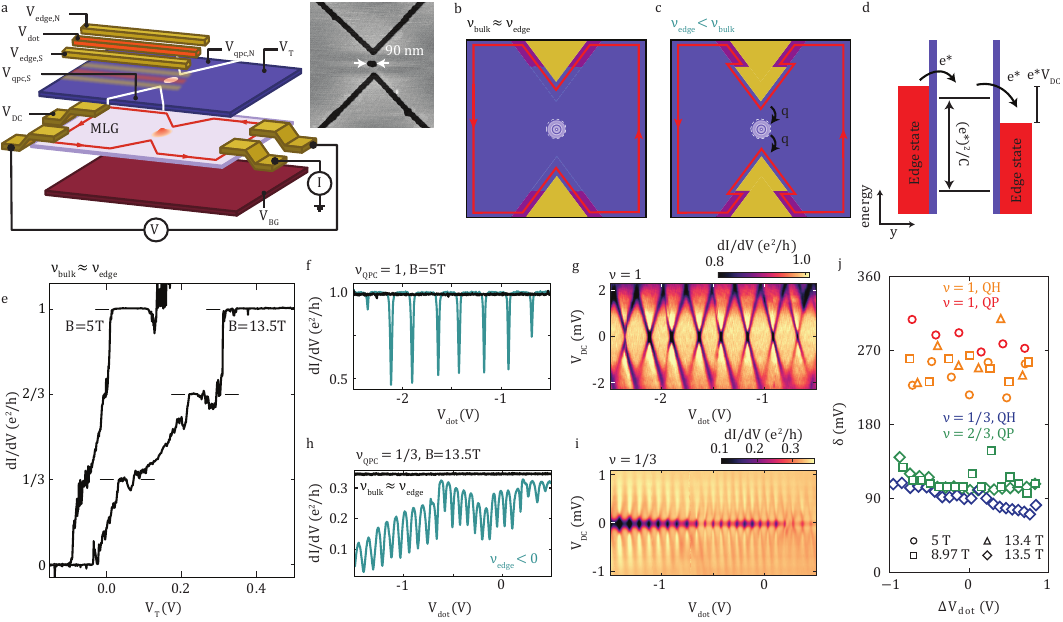} 
    \caption{\textbf{Integer and fractional quantum Hall quantum dot.} 
\textbf{a}, Device schematic.  A monolayer graphene  layer is encapsulated between hexagonal boron nitride dielectrics (not shown) and controlled by 7 field-effect gates in three vertical layers.  A graphite bottom gate (at voltage $V_{\rm B}$) tunes the entire device.  A top graphite gate ($V_{\rm T}$) is patterned using atomic force lithography\cite{cohen_nanoscale_2023} to form a QPC geometry consisting of a central top gate perforated by a hole of 90 nm diameter as well as north and south QPC control gates ($V_{\rm qpc,N}$ and $V_{\rm qpc,S}$) (see atomic force micrograph, inset). Three aluminum finger gates are aligned over the hole ($V_{\rm dot}$) and trenches ($V_{\rm edge,N}$ and $V_{\rm edge,S}$) close to the dot.  
Transport experiments probe the induced DC current $I$ and DC voltage $V$ in response to an applied DC voltage bias $V_{\rm DC}$, as shown. 
\textbf{b}, 
    Edge mode schematic with $N$ and $S$ regions depleted and $\nu_{\rm bulk}\approx \nu_{\rm edge}$, suppressing edge-to-dot tunneling.
    \textbf{c}, Edge mode schematic with $\nu_{\rm edge}<\nu_{\rm bulk}$, enabling edge-to-dot tunneling.
\textbf{d}, Schematic of the dot with edge mode leads. The tunneling charge $e^*$ is the minimal charge supported by the bulk phase: $e$ at $\nu=1$ and $e/3$ at $\nu=1/3$.  
\textbf{e}, $\mathrm{d}I/\mathrm{d}V$ for the regime in panel b at $B = 5\,\rm{T}$ and $13.5\,\rm{T}$.  
 \textbf{f}, $\rm{d}I/\rm{d}V$ as a function of $V_{\rm{dot}}$ for $\nu=1$ and at $B=5\,\rm{T}$ for dot-coupled (teal) and decoupled (black) edge modes.  
 \textbf{g}, Nonlinear response for the teal curve in panel f, showing quasiparticle backscattering Coulomb diamonds with an inferred charging energy of $E_{\rm C}\approx e\times \Delta V_{\rm DC}\approx 2\mathrm{meV}$. 
 \textbf{h}, $\rm{d}I/\rm{d}V$ for $\nu=1/3$ and at $B=13.5\,\rm{T}$ for dot-coupled (teal) and uncoupled (black) edge modes.
\textbf{i},  Nonlinear response for the teal curve in panel h.  The Coulomb diamonds imply a charging energy for fractionally charged quasiparticle backscattering of $E_{\rm C}=\frac{e}{3}\Delta V_{\rm DC}\approx 300\,\mu \mathrm{eV}$. 
 \textbf{j}, $V_{\rm dot}$ spacing of the Coulomb resonances ($\delta$) as a function of the change in $V_{\rm{dot}}$ for several datasets corresponding to quasihole (QH) and quasiparticle (QP) doping at $\nu=1, 1/3,$ and $2/3$ at the indicated magnetic fields.  
} 
\label{fig:Idot} 
\end{figure*}

We consider the possibility of controllably trapping an individual anyon, both as a precision probe of strongly correlated physics and as the fundamental element of a topologically protected quantum information processor.  
FQH liquids are incompressible gapped states of interacting electrons in which the elementary excitations carry fractional charge.  
Because fractionally charged particles are not fundamental, however, the total charge of a finite-sized FQH droplet must remain quantized in units of the whole electron charge.  
To trap a single anyon, then, requires applying a small trapping potential in the two-dimensional bulk of the incompressible fluid, localizing charge at the expense of removing a compensating charge from the sample boundary. 
In principle, the size of this internal quantum dot can be reduced to the point where it contains a single quantized fractional charge (of either sign), providing the simplest instantiation of an anyon as an emergent particle.  This limit resembles a single-electron or single-hole dot in a semiconductor, with the trivial vacuum of the semiconductor replaced by the topologically ordered vacuum of the FQH gap. 

Here we leverage the high sample quality and experimental control available in graphene-based van der Waals heterostructures to study an anyonic quantum dot down to the last fractionally charged quasiparticle.  
This is achieved by bringing the macroscopic edge state---described by a chiral Luttinger liquid (CLL)---into proximity with the dot within a quantum point contact (QPC). 
We then perform charge spectroscopy via tunneling of fractional charges across the topological vacuum from the delocalized outer edge to the localized dot~\cite{jain_quantum_1988,kivelson_methods_1989} and out again. 
In the limit of weak coupling between CLL leads and dot, our system behaves as a single anyon transistor in which transport is governed by the physics of fractional Coulomb blockade.  
Using this Coulomb blockade regime, we demonstrate both trapping of single anyons and quantitative metrology of inter-anyon interactions as a function of fractional dot occupation.  

\section{Tunable integer and fractional dot}
In order to realize a single anyon transistor, our device must satisfy two key requirements: control, and a sufficiently low level of unwanted disorder. 
We achieve the required control using the device geometry shown in \ref{fig:Idot}a, in which a monolayer graphene active layer is controlled by a global graphite bottom gate and a two-layer top gate structure. 
Disorder is minimized by using a graphite top gate built from a graphite flake patterned by anodic oxidation lithography prior to stack assembly\cite{cohen_nanoscale_2023}. 
As shown in the inset, trenches separate the main top gate (held at voltage $V_{\rm T}$) from `North' ($V_{\rm qpc,N}$) and `South' ($V_{\rm qpc,S}$) QPC gates.  A 90nm diameter hole is patterned in the center of the QPC itself, which will define the dot. 
A final aluminum gate layer consists of three finger gates aligned, respectively, over the hole itself to form the `dot' gate ($V_{\rm dot}$) and over the trenches closest to the dot to form the north- and south edge gates ($V_{\rm edge,N}$ and $V_{\rm edge,S}$). 
Four ohmic contacts to the graphene layer allow measurement of current $I$ and voltage $V$ across the QPC in the diagonal conductance geometry.
Details of sample fabrication are described in \ref{fig:device}.  

Our device geometry is conceptually similar to previously explored dot geometries\cite{kataoka_detection_1999,kataoka_kondo_2002, kou_coulomb_2012, mills_localizing_2020,di_luca_quantum_2026,luca_trapping_2026,li_competing_2026} but---crucially---combines a small dot radius with independent tunability of the dot occupation and edge-to-dot tunneling.   To probe the filling factor surrounding the dot, we use $V_{\rm qpc,N}$ and $V_{\rm qpc,S}$ to deplete the south and north regions and force the edge states through the QPC. The resulting conductance reflects the number of transmitted edge modes.  
To tune this transmission further, we use $V_{\rm edge,N}$ and $V_{\rm edge,S}$ to control the edge filling $\nu_{\rm edge}$. When the bulk and edge-gated regions are both set to the same filling $\nu_{\rm edge}\approx \nu_{\rm bulk}=1$, tunneling between the edges and the dot is negligible (\ref{fig:Idot}b).  
Tuning $\nu_{\rm edge}<\nu_{\rm bulk}$ then provides fine experimental control over the dot-edge tunneling (\ref{fig:Idot}c). 
When this coupling is finite, the edge states propagating through the QPC act as reservoirs with tunable tunnel coupling to the discrete energy levels of the dot, as depicted schematically in \ref{fig:Idot}d.  Note that the occupation of the dot may be either positive or negative relative to the filling of the incompressible quantum Hall state within the QPC, and that the tunneling charge $e^*$ is free to take on the minimal quantized fractional value determined by the surrounding bulk of the FQH state. 

Transport through the QPC is shown in \ref{fig:Idot}e for $B=5$ and $13.5~{\rm T}$ for the dot-uncoupled configuration of \ref{fig:Idot}b.  
We observe well quantized plateaus at $\nu=1$ (for $B=5~{\rm T}$) and at $\nu=1/3$ and $2/3$ (for $13.5~{\rm T}$).  
\ref{fig:Idot}f demonstrates the effect of the edge gates. 
With the bulk and edge both set to $\nu=1$, the dot gate does not affect edge mode transmission, which remains at its quantized value (black).  Depleting the edge-gated regions to push the edge modes closer to the dot, we observe sharp, approximately periodic, backscattering resonances marked by dramatic reductions in $\mathrm{d}I/\mathrm{d}V$ (teal). 
\ref{fig:Idot}g shows ${\rm d}I/{\rm d}V$ as a function of $V_{\rm DC}$ and $V_{\rm{dot}}$ in the same regime; the behavior is consistent with resonant backscattering via a dot tuned through a series of Coulomb blockade resonances corresponding to charge degeneracy within the dot.  From the height of the diamonds we extract a charging energy of $e\Delta V_{\rm DC}\approx 2 \rm{meV}$. 
To ensure that the dot is in the intended location, we track the response of the charge degeneracy points as a function of different gate voltages.  As expected, the resonances are nearly insensitive to the QPC gates, very weakly tuned by the edge gates, and very strongly tuned by both the center and bottom gate (see \ref{fig:GateDependence})---consistent with a localized state residing in the saddle point of the QPC.   
The same experiment performed for a 1/3 FQH edge mode is shown in \ref{fig:Idot}h for $B=13.5~{\rm T}$. For  $\nu_{edge}\approx \nu_{bulk}=\nu=1/3$, $V_{\rm dot}$ has no measurable effect on the conductance, which is quantized to $e^2/3h$ (black).  
Upon depleting the edge gated region, we again observe periodic conductance suppression dips (teal) associated with Coulomb diamonds in the nonlinear response (\ref{fig:Idot}i). 
These show a gate-voltage period approximately three times smaller than for $\nu=1$, consistent with addition of charge-$e/3$ quasiparticles to the dot. 
We also observe a strong suppression of the conductance from its quantized value, which is strongest on the Coulomb resonances but is also present in the blockaded regime.  
As shown in \ref{fig:Luttinger}, this is consistent with the leads of the dot being CLLs\cite{wen_chiral_1990,cohen_universal_2023,guerrero-suarez_universal_2025}, which couple via the dot resonances but also either directly or via co-tunneling processes in the blockaded dot. 
Such backscattering is strongly relevant in the renormalization group sense for fractional edge modes\cite{wen_chiral_1990,wen_theory_1992,kane_transport_1992,kane_transmission_1992}, contrasting with the integer case. 

Similar experiments can be performed at different magnetic fields and for different filling factors, including $\nu=2/3$. 
\ref{fig:Idot}j shows the gate voltage spacing $\delta=V_{\rm{dot}}^{N+1,N}-V_{\rm{dot}}^{N,N-1}$ (where $V_{\rm{dot}}^{N+1,N}$ corresponds to the dot gate voltage at the $N_{qp}\rightarrow N_{qp}+1$ transition). 
For the integer quantum Hall state, $\delta\approx 250-300 \rm{mV}$, while at both fillings 1/3 and 2/3 $\delta$ is $80-100\,\rm{mV}$, or approximately three times smaller than the integer case.
This can be understood within a simple electrostatic model of the energy of a dot with $N_{\rm qp}$ quasi-particles, $E = \frac{{e^*}^2 N_{\rm qp}^2}{2 C_{\rm d}} - \alpha e^* V_{\rm dot} N_{\rm qp}$. Here $C_{\rm d}$ is the capacitance between the dot and its surroundings, and $\alpha$ is the dimensionless ``lever arm'' between the dot gate and the dot. Assuming these phenomenological parameters are determined primarily by geometry, and therefore ignoring any dependence on $N_{\rm qp}$ or $\nu$, minimizing $E$ predicts a voltage spacing $\delta = \frac{1}{\alpha}\frac{e^*}{C_{\rm d}}$  proportional to the quasiparticle charge.  
Under these assumptions, our results are consistent with addition of charge $e/3$ particles to the dot at $\nu=1/3$ and $2/3$, in contrast to experiments on larger, higher occupation mesoscopic structures\cite{kou_coulomb_2012,di_luca_quantum_2026,bid_shot_2009,kim_aharonovbohm_2026} which have reported $2e/3$ charge at $\nu=2/3$. 
However, we note that our simple phenomenological model ignores possible variations in the orbital structure or importance of  interparticle interactions within the dot, which are explored below.

\begin{figure*}[t]
    \centering
    \includegraphics[]{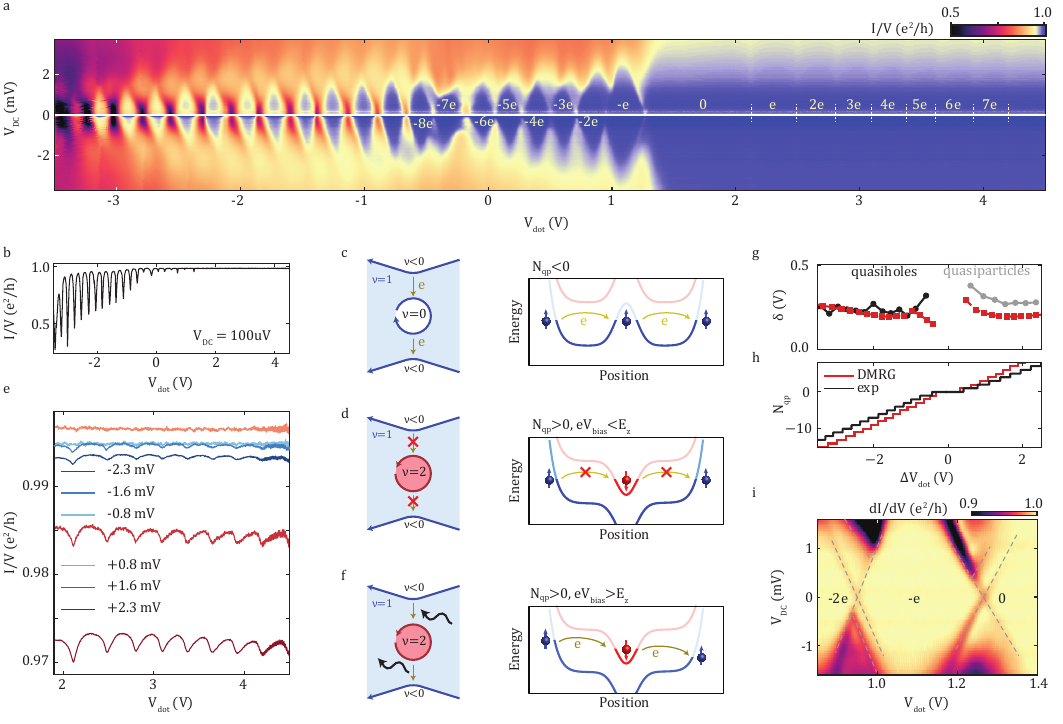} 
    \caption{
    \textbf{Single quasihole-quasiparticle control in the integer regime.} 
    \textbf{a}, $I/V$ as a function of $V_{\rm DC}$ and $V_{\rm dot}$ under the same conditions as in \ref{fig:Idot}g, but with $V_{\rm edge}=-3.5~\mathrm{V}$. \textbf{b}, Low-bias data acquired under the same conditions as in \textbf{a}. 
    \textbf{c}, Schematic of the dot embedded in the QPC for $N_{\rm qp}<0$ and 
   \textbf{d}, $N_{\rm qp}>0$.  The left panels show a real-space diagram while the right panels show the energy as a function of y-coordinate across the QPC neck.  Edge-to-dot tunneling is allowed for $N_{\rm qp}<0$ but spin-blockaded for $N_{\rm qp}>0$.
    \textbf{e}, High-bias data acquired under the same conditions as panel \textbf{a} for $N_{\rm qp}>0$.
    \textbf{f},  Schematic of the dot embedded in the QPC for $N_{\rm qp}>0$ and applied bias exceeding the Zeeman energy. Magnon emission and absorption enable spin-nonconserving tunneling between the edge state and the dot, lifting the spin blockade. 
    \textbf{g}, Comparison of the resonance spacing $\delta$ for quasiholes and quasiparticles superimposed on the DMRG calculations. \textbf{h}, Comparison between theory and experiment for the dot charge occupancy $N_{\rm qp}$. \textbf{i}, Zoomed-in view of \textbf{a} around the few quasiholes regime, highlighting the presence of excited states above 300 $\mu$V.}     
    \label{fig:dot_anti} 
\end{figure*}

\section{Quantum Hall spin blockade}
A key question concerns the absolute occupation of the dot: does the observed Coulomb blockade  correspond to a dot of quasiparticle or quasiholes---and can the dot be emptied completely? 
To address this, we study the $\nu=1$ state where the near-perfect spin polarization of the quantum Hall ferromagnet provides a sharp experimental signature of the dot polarity. 
\ref{fig:dot_anti}a shows the conductance $I/V$ over an extended $V_{\rm dot}$ range at $\nu=1$. 
For $V_{\rm dot}<1.5~{\rm V}$, we observe well-defined Coulomb diamonds; in contrast, no features are visible near zero bias for $V_{\rm dot}>1.5~{\rm V}$ (see also \ref{fig:dot_anti}b). 
We associate the region of observable zero-bias Coulomb blockade as the regime in which the dot traps quasiholes; in this case, the edge state and quasiholes occupying the dot have the same spin (\ref{fig:dot_anti}c), allowing quasiholes to tunnel from edge to dot.  
Within this picture, we may assign the charge on the  dot by counting from the last observed resonance near $V_{\rm{dot}}\approx 1.25V$.  Quasielectron states, in contrast, are drawn from the opposite spin excited Landau level of the $\nu=1$ quantum Hall ferromagnet.  As illustrated in \ref{fig:dot_anti}d, spin conservation forbids tunneling, in a quantum Hall ferromagnetic version of spin blockade.  

Close examination of the finite bias behavior, however, reveals periodically spaced conductance dips emerging for $|V_{\rm DC}|\gtrsim1~{\rm mV}$ 
with similar periodicity to the Coulomb resonances observed for quasiholes (see \ref{fig:dot_anti}e).  We attribute this effect to inelastic processes in which tunneling from the edge into the dot is accompanied by the absorption of magnons. 
As shown previously\cite{wei_electrical_2018,
zhou_solids_2019,zhou_strong-magnetic-field_2022,assouline_excitonic_2021}, neutral spin waves can be excited in the insulating bulk of the $\nu=1$ quantum Hall ferromagnet when a potential difference greater than the Zeeman energy, $E_{\rm Z}\approx 0.58 \,\mathrm{meV}$ at $B=5\,\mathrm{T}$, is induced by a finite bias between edge modes of opposite spin (which are generically located near contacts, in addition to the gate-defined dot).  Tunneling between the edge mode and opposite spin dot is then allowed by the absorption of a magnon, as shown in \ref{fig:dot_anti}f.  

Armed with this understanding, we may assign the quasielectron occupations as well (indicated on \ref{fig:dot_anti}a). The experimentally determined charge transition voltages may then be compared to numerical simulations (see \ref{fig:dot_anti}g--h). 
In these simulations, we first compute the applied electrical potential via finite-element electrostatic simulations of the heterostructure.  
Following recent scanning tunneling microscopy measurements in similar structures\cite{yu_visualizing_2025}, we also allow for a  work-function shift  localized on the vertical face of the anodically oxidized trench; the value of this shift is fixed from the observed value of $V_{\rm dot}$ corresponding to $N_{\rm qp}=0$.  
We then simulate the ground state of the dot using the density matrix renormalization group (DMRG) in which the device is modeled as a finite sphere with gate-defined dot potential at one pole. 
Our simulations account for the screening due to the filled Landau levels within the random phase approximation, but neglect the effect of fringe electric fields from the edge and QPC gates and approximate the dot geometry as perfectly circular (see Methods).  
The numerical data reproduce 
the observed experimental trend in which $\delta$ 
decreases at large $|N_{\textrm{\rm qp}}|$, 
the observed width of the intrinsic $\nu=1$ ferromagnetic gap separating $N_{\textrm{\rm qp}}=\pm1$, and (at least qualitatively) the particle-hole asymmetry which arises in the model from the finite work function of the anodic oxidation trench (see \ref{fig:dot_potential}). 

Our realization of a single electron dot also allows us to quantify several key figures of merit for realizing proposed topological qubits in a van der Waals platform\cite{bonderson_scalable_2023}. 
First, qubits generally require the spectrum of excited states to be well separated energetically from the ground state. This can be challenging in a Landau level, where the single-particle dispersion is flat and single-particle level spacings are consequently generated solely by the curvature of the confining potential on the scale of the magnetic length~\cite{yang_addition_1993}. 
\ref{fig:dot_anti}i shows a high-resolution detail of $\mathrm{d}I/\mathrm{d}V$ measured between dot occupation 0 and -2e.  Several excited-state resonances can be identified (marked by the dashed lines), appearing as additional enhanced backscattering features running approximately parallel to the diamond edges. 
We extract excitation energies in the range of $\Delta \approx 0.3-0.8~\mathrm{meV}$, well above typical measurement temperatures. 
A second figure of merit is the low-frequency charge noise, which will impact fidelity of topologically unprotected gates. 
 The charge-noise spectral density from $10^{-4}$ to $10^{-2}$ Hz extracted from the time dependence of several Coulomb peak positions follows a $1/f^\alpha$ behavior, with a noise amplitude of $\sqrt{S_{\rm q}}\approx 10^{-3} e/\sqrt{\mathrm{Hz}}$ at 10 mHz (see \ref{fig:extended_noise} and \ref{tab:charge_noise}). 
For comparison, measurements in aluminum single electron transistors show similar noise at frequencies two orders of magnitude higher\cite{zorin_background_1996}.  
In terms of the gate-referred charge noise commonly reported in semiconductor quantum dots, our graphene dot is among the best reported in a solid state system\cite{stehouwer_exploiting_2025,kranz_exploiting_2020,paquelet_wuetz_reducing_2023,basset_evaluating_2014,dial_charge_2013,richter_low-frequency_2026, huang_enhancing_2026}.  

\section{Fractional dot in the single anyon regime}

Realizing a single anyon dot in the FQH regime presents several experimental challenges as compared to the integer case.  No qualitative spin blockade signature of the $N_{qp}=0$ state is expected
because both the ground state and low-lying excitations at $\nu=1/3$ have the same spin polarization~\cite{polshyn_quantitative_2018}.
Determining $N_{\rm qp}$ must, as a consequence, be done via quantitative electrostatic calibration. 
The tunneling charge itself can also differ depending on how the dot is embedded in the QPC. In the desired limit, the dot mediates weak backscattering between top and bottom chiral Luttinger liquid edge modes, which are otherwise fully transmitted when the dot is in Coulomb blockade. 
In this case, the dot may freely adjust its charge by any multiple of $e/3$, and charge degeneracy points manifest as conductance dips due to the tunneling of $e/3$ charges through the dot. 
When the QPC is close to pinch off, in contrast, the dot mediates forward scattering of whole electrons and the dot charge remains quantized in units of $e$.  
Experimentally, which regime is realized at a given $N_{\textrm{\rm qp}}$ can be sensitive to all of the gate voltages in addition to the temperature and bias.

\begin{figure*}[t]
    \centering
    \includegraphics[width=\textwidth]{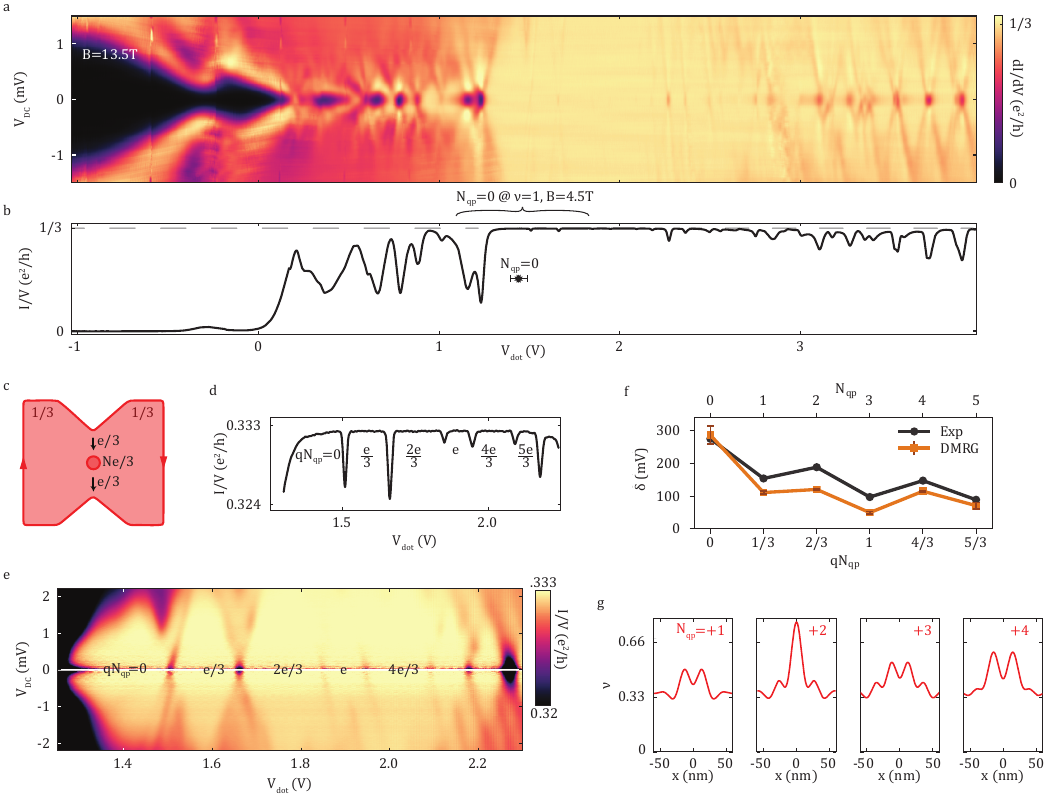} 
    \caption{\textbf{Single anyon dot.} \textbf{a}, $\rm{d}I/\rm{d}V$ as a function of $V_{\rm DC}$ and $V_{\rm dot}$, measured at $\nu_{\rm bulk}=1/3$. \textbf{b}, I/V with $V_{\rm DC}=100 \mu\mathrm{V}$. The bracket indicates the span of the $N_{\rm qp}=0$ diamond for $\nu=1$ at $B=4.5\,T$. The range shows the expected position of the $N_{\rm qp}=0$; the range represents an error bar derived from our calibration of quantum capacitance effects.  
    \textbf{c}, 
    Schematic of backscattering via fractional Coulomb blockade of the dot.  \textbf{d}, Detail of the conductance for low $N_{\rm qp}>0$, with the charge sectors indicated.   
    \textbf{e},  
    High-resolution $I/V$ in the few quasiparticle regime. 
    \textbf{f}, Spacing $\delta$ in the few-quasiparticle regime, measured experimentally (black) and calculated using DMRG (orange). 
    \textbf{g}, Local filling-factor profiles obtained using DMRG for the first few quasiparticles.
    }
    \label{fig:fraction} 
\end{figure*}

Our multigate geometry provides sufficient control to fix both $N_{\rm qp}$ and the tunneling charge separately.   
To calibrate $N_{\rm qp}$, we first determine the gate voltage range corresponding to $N_{\textrm{\rm qp}}=0$ at $\nu=1$ and $B=4.5\,\mathrm{T}$ and then triple the applied magnetic field. Neglecting magnetic field dependent chemical potential shifts ('quantum capacitance') of the graphene and graphite gate layers, the range of gate voltages corresponding to $N_{\rm qp}=0$ at $\nu=1/3$ and $B=13.5\mathrm{T}$ should coincide with that for $\nu=1$ and $B=4.5\,\mathrm{T}$.  
As shown in \ref{fig:zeroqp_identification}, these quantum capacitance corrections are negligible in our device, amounting to a shift equivalent to less than one fractionally charged quasiparticle in our dot. With $N_{\rm qp}=0$ at fractional filling fixed, we may then adjust the QPC electrostatics to ensure that these dot states mediate only weak backscattering between the CLL edge modes passing through the QPC. 
As described in \ref{fig:virtual}, optimization over top, bottom, QPC, and edge gate voltages is sufficient to tune the dot-mediated tunneling regime for fixed $N_{\rm qp}$ across a broad range.  For example, for $N_{\rm qp}\ll-1$, we observe a crossover between forward scattering of whole electrons and backscattering of fractional charges. For $N_{\rm qp}\gg1$, we observe a crossover between backscattering of fractional charges from a $1/3$ edge in the QPC to forward scattering of fractional charges from an internal $1/3--2/5$ boundary edge mode. 
Most importantly, however, this tunability allows us to achieve a finite but small backscattering coupling in the low quasiparticle regime.  

\ref{fig:fraction}a--b shows the result of this optimization.  
The nontrivial coupling between $V_{\rm dot}$ and \textit{both} $N_{\rm qp}$ and the pinch-off behavior of the QPC is evident in the overall scale of the QPC transmission, which is tuned from zero to near $e^2/(3h)$; resonances associated with the dot are observed across most of this range. 
The bracket in \ref{fig:fraction}b marks the extent of the $N_{\rm qp}=0$ charge sector at $\nu=1$ and $B=4.5~\mathrm{T}$; the center of this region, accounting for the measured quantum capacitance and the possible error in that calibration, is also shown.  
This comparison indicates that the few quasiparticle regime indeed lies in the regime of weak backscattering, as depicted schematically in \ref{fig:fraction}c, with the large diamond centered at $V_{\mathrm{dot}}=1.44~\mathrm{V}$ corresponding to $N_{\textrm{\rm qp}}=0$.  

\ref{fig:fraction}d--e show a high-resolution view of the few quasiparticle regime.  
We were unsuccessful in accessing the few quasihole regime in weak backscattering: the QPC either pinches off (as in \ref{fig:fraction}b) or the tunnel coupling between dot and edge modes is too small to resolve the backscattering current. However, our data do allow us to analyze the addition energies of the first few anyonic quasiparticles. These are irregularly spaced when compared to the highly regular behavior in \ref{fig:Idot}h--j, which corresponds to quasihole doping of the dot with $N_{\rm qp}\ll-1$.  
To make sense of this observation, we make use of the fact that the small number of quasiparticles makes this a tractable problem for numerical simulations even in the fractional regime. 
\ref{fig:fraction}f compares the  experimentally measured $\delta$ with the numerical results for the variational ground states in the different charge sectors (\ref{fig:fraction}g).  The calculation reproduces the width of the incompressible $N_{\rm qp}=0$ sector, the 100--200 $\mu V$ scale of the charge stability sectors, and the alternating ``even-odd'' effect characterizing the stability of the first few quasiparticle states for $N_{\rm qp}$ up to 5.  Within the DMRG model, the observed irregularity of the addition energy arises from a competition between spatially extended states that minimize the occupation of higher-energy $\Lambda$-level\cite{jain_composite-fermion_1989} dot orbitals and spatially compact states that minimize the confinement energy (see \ref{fig:cf_filling}).   The striking even-odd effect observed in this regime is thus an accidental effect arising from the orbital structure of the dot.

\begin{figure*}[t]
    \centering    \includegraphics[width=\textwidth]{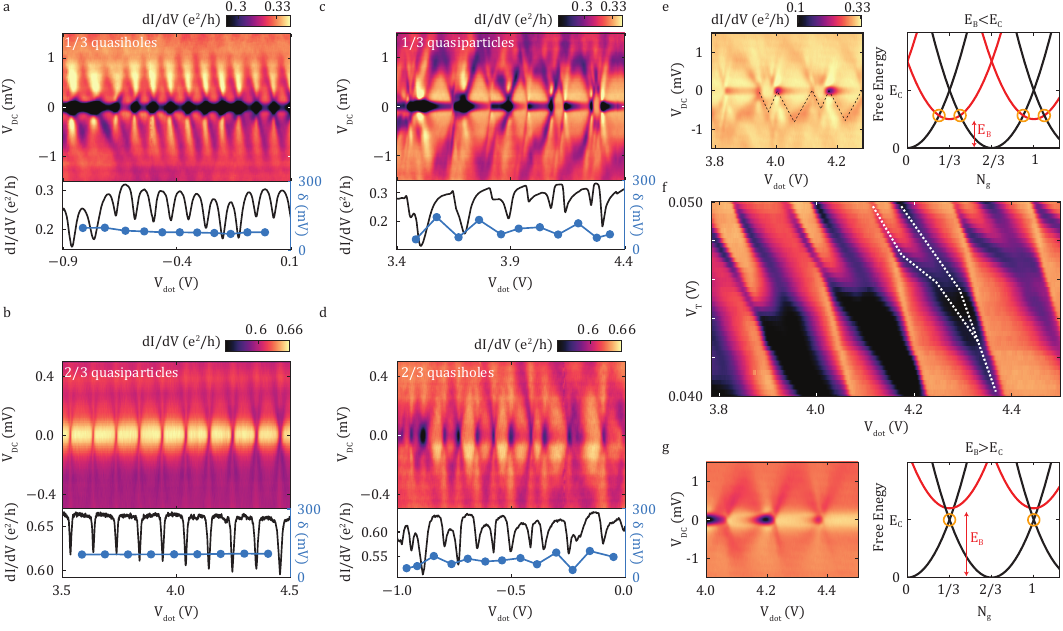} 
    \caption{\textbf{Evidence for anyon attraction.}  Upper panels of \textbf{a--d}, $\mathrm{d}I/\mathrm{d}V$ as a function of $V_{\rm DC}$ and $V_{\rm dot}$ for dot filling factors $\nu_{\rm dot}=1/3$ (\textbf{a}, \textbf{c}) and $\nu_{\rm dot}=2/3$ (\textbf{b}, \textbf{d}). Panels \textbf{a} and \textbf{d} correspond to the quasihole regime, whereas \textbf{b} and \textbf{c} correspond to the quasiparticle regime. The lower panels show linecuts taken at $V_{\rm DC}=0$ of the corresponding upper panels (black traces) and the relative $\delta$ spacings (blue trace). Panels \textbf{a} and \textbf{b} exhibit regularly spaced Coulomb diamonds and resonances, whereas \textbf{c} and \textbf{d} exhibit irregular spacings and an alternation between small and large diamonds, indicative of an even--odd effect. \textbf{e}, Left panel, Zoomed-in view of the even–odd effect in the $\nu=1/3$ quasiparticle regime at $V_{\rm T}=0.05\,\mathrm{V}$, showing alternating small and large Coulomb diamonds. Right panel, free-energy parabolas $F$ as a function of gate-induced excess charge $N_{\rm g}$. States with odd quasiparticle occupation are shifted upward by the binding energy $E_{\rm B}$, with $E_{\rm B}<E_{\rm C}$. \textbf{f}, Conductance $I/V$ measured at $V_{\mathrm{DC}}=55 \mu\mathrm{V}$ as a function of $V_{\mathrm{dot}}$ and $V_{\rm T}$ in the $\nu=1/3$ quasiparticle regime. As $V_{\rm T}$ decreases, neighboring charge-degeneracy lines merge, evolving from the alternating spacings observed in \textbf{e} to a charging periodicity corresponding to $2e/3$. \textbf{g}, Left panel, same as left panel of \textbf{e}, but at $V_{\rm T}=0.04 \mathrm{V}$, where the intervening charge sector is absent, consistent with the addition of $2e/3$ quasiparticle molecules to the dot. Right panel, free-energy description for $E_{\rm B}>E_{\rm C}$, corresponding to the lower part of \textbf{f}.}
    \label{fig:molecular} 
\end{figure*}

\section{Evidence for anyon binding}

Motivated both by the possibility of intrinsic anyon superconductivity\cite{laughlin_superconducting_1988,lee_anyon_1989,shi_superconductivity_2026} as well as 
apparent pairing in large quasiparticle ensembles in the FQH regime~\cite{ghosh_coherent_2025,samuelson_slow_2026,kim_aharonovbohm_2026,henzinger_controlled_2026}, 
recent theoretical work has considered whether anyons may form bound `molecules' with a charge larger than the elementary one~\cite{xu_dynamics_2025,gattu_molecular_2025,li_bound_2026,wang_anyon_2026}.
Such binding arises due to the real-space structure of the anyonic wavefunctions, which feature regions of both positive and negative charge relative to the uniform background.  
A sharp criterion for binding within a quantum dot can be defined by considering the total interaction energy $E_{\textrm{int}}(N)$ for an $N$-quasiparticle state (e.g. the energy after removing the applied single-particle potential).  
Intrinsic binding can then be defined as a non-convex $E_{\textrm{int}}(N_{qp})$. 
When the applied potential is spatially uniform, the convexity condition is equivalent to the usual definition of the binding energy: the difference in energy between well-separated quasiparticles and  their minimum-energy configuration.  
For a dot, minimizing the energy in the presence of an applied potential ($N(N_{qp}) = \textrm{argmin}_{N_{qp}} (E_{\textrm{int}}(N_{qp})  - \alpha N e^* V)$) leads, for a non-convex interaction energy, to the complete suppression of some charge sectors---the hallmark of attraction and key experimental phenomenology associated with intrinsic binding.  
This `missing charge sector' effect is well studied for electrons, notably in superconducting single electron transistors\cite{tuominen_experimental_1992,hekking_coulomb_1993} and more recently in a carbon nanotube device where a dielectric medium provides the source of the attractive interaction~\cite{hamo_electron_2016}.  

Our DMRG calculations predict that $\pm e/3$ anyons do not bind in the absence of confinement for the parameters of the current experiment and that the interaction energy in the presence of the confining dot potential remains convex at low $N_{qp}$.  However, they do predict binding of anyons in a uniform system at $\nu=2/5$ (see Methods, \ref{tab:binding}).  This motivates a search for intrinsic binding at large $N_{qp}$, where the orbital structure of the dot is expected to include contributions from similar composite fermion orbitals as those characterizing the ground state at higher filling.  
\ref{fig:molecular}a--d show maps of the addition spectra for quasiparticles and quasiholes at both $\nu=1/3$ and $\nu=2/3$ for $|N_{\rm qp}|\approx \text{10-20}$.  We observe a striking symmetry: quasiholes at $\nu=1/3$ and quasiparticles at $\nu=2/3$ exhibit highly \textit{regular} charging energies and peak spacings (\ref{fig:molecular}a,b), while quasiparticles at $\nu=1/3$ and quasiholes at $\nu=2/3$ display strongly \textit{irregular} addition spectra (\ref{fig:molecular}c,d). 
This is consistent with a particle-hole symmetry of the singly-degenerate Landau level, which maps quasiholes at $\nu=1/3$ to quasiparticles at $\nu=2/3$.  Neglecting Landau level mixing, the Coulomb interaction respects this symmetry whereas the edge-confinement and disorder potentials do not.
We attribute the contrasting experimental phenomenology to the difference in occupied orbitals, which involve only the lowest $\Lambda$ level for 1/3 quasiholes (and 2/3 quasiparticles) but higher $\Lambda$-levels for 1/3 quasiparticles (and 2/3 quasiholes).  
The qualitative distinction between regular and irregular addition spectra persists over a wide range of gate configurations (\ref{fig:Spacing_Regular}, \ref{fig:Spacing_Molecule} and \ref{fig:Molecular_wideRange}).

Close examination of the behavior at high $N_{qp}$ reveals evidence for anyon binding in the dot. 
\ref{fig:molecular}e focuses on a narrow range of gate voltage in which we observed a particularly pronounced even-odd effect characterized by an alternating pattern in the stability of the quasiparticle charge sectors. 
We then study the evolution of the charging peaks as a function of the electrostatic confinement, tracking individual charging peaks as a function of $V_{\rm T}$ (\ref{fig:molecular}f). As $V_{\rm T}$ is lowered, the weak charge sectors disappear; dot occupation now transitions directly between stable charge sectors $N_{\rm qp} \to N_{\rm qp}+2$ (\ref{fig:molecular}g). These observations are easily understood in a model that includes an attractive interaction, depicted schematically in  free energy diagrams of \ref{fig:molecular}e and g.  The even-odd regime corresponds to a binding energy comparable to, but less than the classical charging energy, while the `molecular' regime arises when $E_{\rm B}$ exceeds $E_{\rm C}$. 

We note that the disappearance of a charge sector places a finite lower bound on the strength of attraction: it implies $E_{\rm B}>E_{\rm C}$, while intrinsic pairing in a uniform state arises for any positive $E_{\rm B}$.  However, binding in our dot, which is embedded within a $\nu=1/3$ vacuum, does not imply that intrinsic binding occurs at 1/3.  Instead, one possibility is that the dot localizes a puddle of $\nu = 2/5$, and charge enters as $m e / 5$ anyonic molecules together with compensating charges at the outer boundary of the dot to ensure a total change in charge of  $2 e / 3$. 
Alternatively, the dot may form a uniform droplet of $\nu=2/3$ surrounded by the $\nu = 1/3$ bulk.  If charge enters via the expansion of the $2/3 - 1/3$ edge state, the pairing could be interpreted as a property of the $2/3 - 1/3$  edge. 

\section{Discussion}

Our observation of anyon binding has immediate implications for the interpretation of experiments in the FQH regime. Anyon binding may account for the anomalous charge observed in transport~\cite{samuelson_slow_2026, di_luca_quantum_2026, luca_trapping_2026,kim_aharonovbohm_2026,henzinger_controlled_2026,kim_selective_2026}  and shot noise~\cite{ghosh_coherent_2025,bid_shot_2009} experiments. Although the wave functions of our dot at large $N_{qp}$ are not known precisely, our numerics at small $N_{qp}$ suggest it rapidly adopts a `wedding cake' structure~\cite{gutierrez_interaction-driven_2018} of nested edge modes separated by incompressible rings.  
Further experiments may clarify the precise relation between this binding and the pairing-like phenomenologies observed in other mesoscopic samples.  Certainly, our experiments leave no doubt that the subtle energetics of `anyon chemistry' must be accounted for in any experiment performed at large anyon number.  

Of course, the nonuniversal energetic considerations underlying anyon chemistry become irrelevant when the quasiparticle number is reduced to 1. 
Our realization of single-anyon dots thus represents a key advance towards precision physics of anyons, and the elucidation of their remaining properties.  
We expect that the same experimental techniques will enable the confinement of a single nonabelian anyon, where the charging energy should be approximately half of that for the e/3 quasiparticles studied here. 
The exceptional charge stability we achieve in the current geometry, attributed to the strong confinement and oxide-free dielectric environment, also bodes well for demonstrating a topological qubit, where stability of charge below a finite threshold is necessary. 
Examining the phase coherence of charge qubits in the integer quantum Hall regime may provide a further benchmark of the pristine nature of the dielectric environment, as recently demonstrated for electrons on solid neon\cite{zhou_electron_2024}. 
More immediately, our geometry can be directly adapted to create a fractional charge pump\cite{simon_proposal_2000}, providing metrology-grade measurements of the quantized fractional charge.  

\textbf{Acknowledgments.}
We acknowledge discussions with Dima Feldman, Ruben Seoane Souto, Maksim Borovkov, Erez Berg, Steve Kivelson, Bert Halperin, and Owen Sheekey.  
Work at UCSB was primarily supported by the Office of Naval Research under award N00014-23-1-2066 and by the Brown Science Foundation under a Brown Investigator Award to AFY. KW and TT acknowledge support from the CREST (JPMJCR24A5), JST and World Premier International Research Center Initiative (WPI), MEXT, Japan. 
MPZ and ASM were supported by the U.S. Department of Energy, Office of Science, Office of Basic Energy Sciences, Materials Sciences and Engineering Division, under Contract No. DE-AC02-05CH11231, within the van der Waals Heterostructures Program (KCWF16).

\textbf{Author Contributions.}
M.V. fabricated the devices with the support of C.L. and supervision of A.F.Y. 
M.V. and W.W. performed the measurements with the help of N.S. and supervision of A.F.Y. 
M.Z. and L.C. developed the measurement electronics.
A.S.M. and J.F. performed the theoretical analysis under the supervision of M.P.Z. T.W. developed the DMRG code for the Haldane sphere. 
T.T. and K.W. grew the hBN crystals.
M.V., A.F.Y., M.P.Z., W.W. and A.S.M. wrote the manuscript with input from all authors.

\setcounter{equation}{0}
\setcounter{figure}{0}
\setcounter{table}{0}
\setcounter{section}{0}
\makeatletter
\renewcommand{\theequation}{S\arabic{equation}}
\renewcommand{\thefigure}{\arabic{figure}}
\renewcommand{\figurename}{Extended Data Fig.}
\def\p@figure{Extended Data Fig.~}
\makeatother

\section{Methods}

\subsection{Stack preparation}
The heterostructure was assembled using a standard van der Waals dry-transfer technique. Graphite and hexagonal boron nitride (hBN) flakes were mechanically exfoliated from bulk crystals onto low-resistivity Si substrates covered with thermally grown SiO$_2$. Substrates with oxide thicknesses of 285~nm and 90~nm were used for hBN and graphite, respectively. Before exfoliation, the substrates were treated in a plasma asher at 50~W and 300~mTorr to promote flake adhesion. Suitable flakes were identified using optical microscopy and subsequently characterized by atomic force microscopy to assess their thickness, surface cleanliness, and the presence of residues or other impurities.\\
The heterostructure was assembled using a polydimethylsiloxane (PDMS) dome mounted on a transfer slide and covered with two layers of homemade poly(bisphenol A carbonate) (PC). The graphite gates, graphene monolayers, and hBN flakes were picked up at temperatures of 90~$^{\circ}$C, 65~$^{\circ}$C, and 110~$^{\circ}$C, respectively.\\
The completed stack was released onto a doped Si substrate covered with a 285-nm-thick layer of thermally grown SiO$_2$ by heating the transfer polymer to approximately 160~$^{\circ}$C with prepatterned contact pads and electron beam alignment markers.
The hBN thicknesses are $50~{\rm nm}$ between the bottom graphite gate and the monolayer graphene, $32~{\rm nm}$ between the monolayer graphene and the aligned top gates, and $35~{\rm nm}$ between the aligned top gates and the top graphite gate (\ref{fig:device}a--e).

\subsection{Device fabrication}

\begin{figure*}
\centering
\includegraphics[]{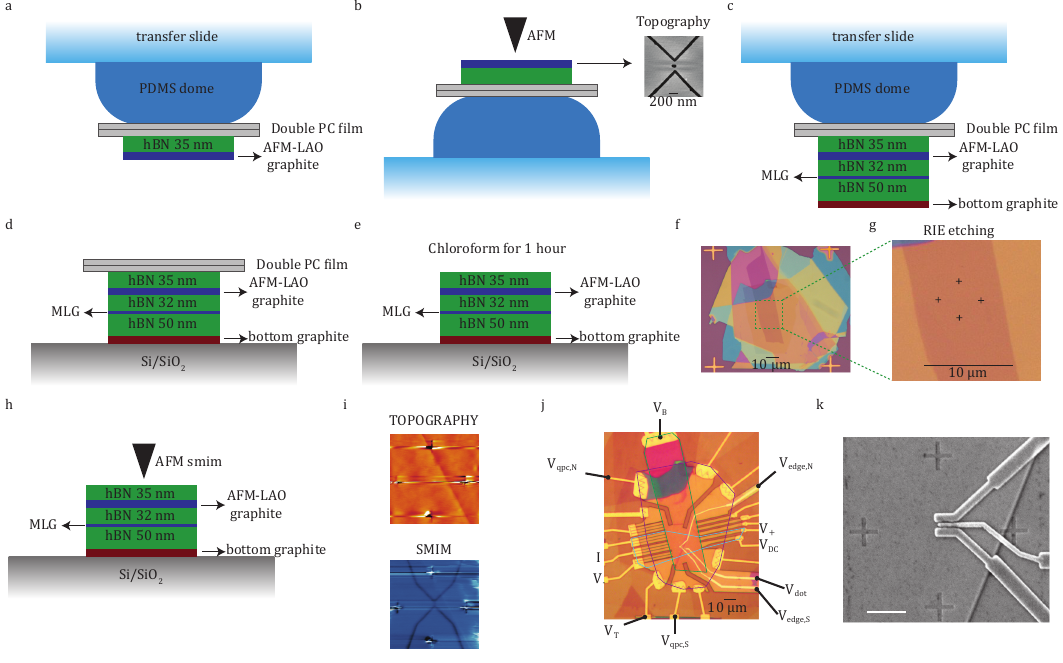} 
\caption{\textbf{Device fabrication.}
\textbf{a,} Schematic of a VdW stack on the underside of a PDMS dome after pickup of the graphite top gate, which was prepatterned using AFM-LAO.
\textbf{b,} AFM image acquired on the transfer-slide to assess the quality of the pre-patterned AFM-LAO gates.
\textbf{c,} Schematic of the completed VdW stack.
\textbf{d,} Stack deposited on an Si/SiO$_2$ substrate.
\textbf{e,} Removal of the PC film.
\textbf{f,} Optical micrograph of the device after the stack was deposited on an Si/SiO$_2$ substrate.
\textbf{g,} RIE etching for defining shallow cross markers in the proximity of the pre-patterned AFM-LAO gate.
\textbf{h,} After the cross markers were etched, an AFM image was acquired for the subsequent alignment step.
\textbf{i,} Top panel: AFM topography of the completed van der Waals stack. The crosses are alignment markers etched into the hBN to a depth of approximately 10 nm. The AFM-LAO-defined graphite gates are not visible in the topography because they are buried beneath the 35-nm-thick hBN layer. Lower panel: A scanning microwave impedance microscopy (sMIM) image acquired simultaneously with the AFM topography shown in the top panel. Both the etched alignment markers and the buried AFM-patterned graphite gates are clearly visible, enabling precise alignment of the metallic gates with respect to the quantum dot and the cut regions in the north and south gates.
\textbf{j,} Optical micrograph of the device after completion of all nanofabrication processes. The cyan, purple, and green contours outline the monolayer graphene, and graphite top- and bottom gates, respectively.
\textbf{k,} SEM micrograph acquired after deposition of the metallic finger gates, confirming their correct positioning relative to the etched markers. The white scale bar corresponds to $1 \mu$m.}
\label{fig:device} 
\end{figure*}

\begin{figure*}
\centering
\includegraphics[]{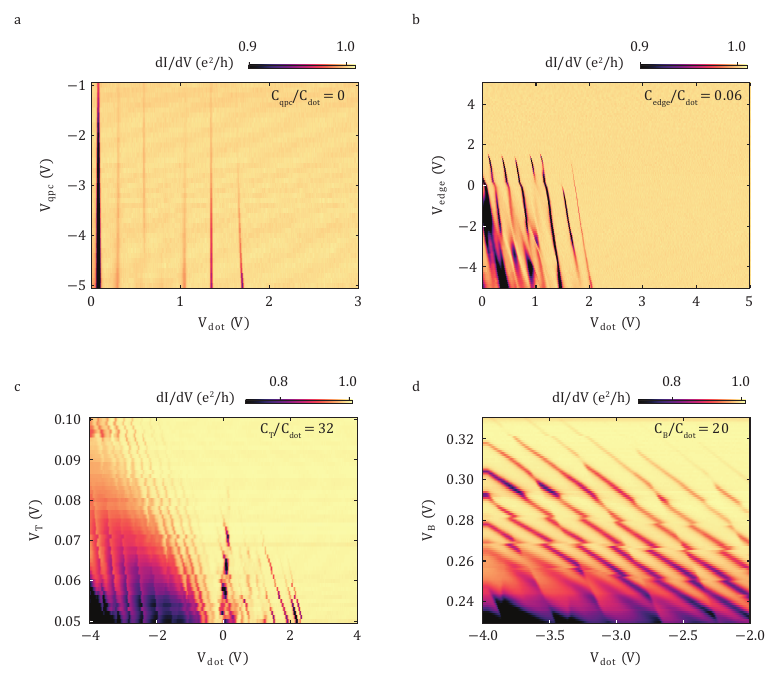} 
\caption{\textbf{Gate dependence of the integer dot.}
\textbf{a}, Zero-bias $\mathrm{d}I/\mathrm{d}V$ as a function of $V_{\rm qpc}=V_{\rm qpc,N}=V_{\rm qpc,S}$ 
and $V_{\rm dot}$, measured in the $\nu=+1$ regime at $B=5~\mathrm{T}$, $V_{\rm B}=0.3~\mathrm{V}$, and $V_{\rm T}=0.07~\mathrm{V}$ [$V_{\rm T}=0.04~\mathrm{V}$]. The edge-gate voltage is fixed at $V_{\rm edge}=-1.5~\mathrm{V}$ 
\textbf{b}, The same with $V_{\rm edge}=V_{\rm edge,N}=V_{\rm edge,S}$ and the QPC-gate voltage relation fixed at $V_{\rm qpc}=-3~\mathrm{V}$.
\textbf{c}, Zero-bias $\mathrm{d}I/\mathrm{d}V$ as a function of $V_{\rm T}$ and $V_{\rm dot}$, measured at $V_{\rm B}=0.3~\mathrm{V}$, $V_{\rm edge}=-5~\mathrm{V}$, $V_{\rm qpc}=-3~\mathrm{V}$, and $B=5~\mathrm{T}$.
\textbf{d}, Zero-bias $\mathrm{d}I/\mathrm{d}V$ as a function of $V_{\rm B}$ and $V_{\rm dot}$, measured at $V_{\rm T}=0.1~\mathrm{V}$, $V_{\rm edge}=-3~\mathrm{V}$, $V_{\rm qpc}=-5~\mathrm{V}$, and $B=5~\mathrm{T}$. In each panel, the capacitance ratio of the gate plotted on the vertical axis to the dot gate is indicated. Variations in $V_{\rm qpc}$ and $V_{\rm edge}$ have little effect on the resonance positions, consistent with a dot confined by the saddle-point potential surrounding the etched hole. Nevertheless, $V_{\rm edge}$ remains weakly coupled to the dot, with a lever arm approximately one-fifteenth that of $V_{\rm dot}$. The top and bottom gates couple more strongly to the dot than $V_{\rm dot}$ because of their closer proximity to the monolayer graphene.}
\label{fig:GateDependence} 
\end{figure*}

\begin{figure*}
\centering
\includegraphics[width=6in]{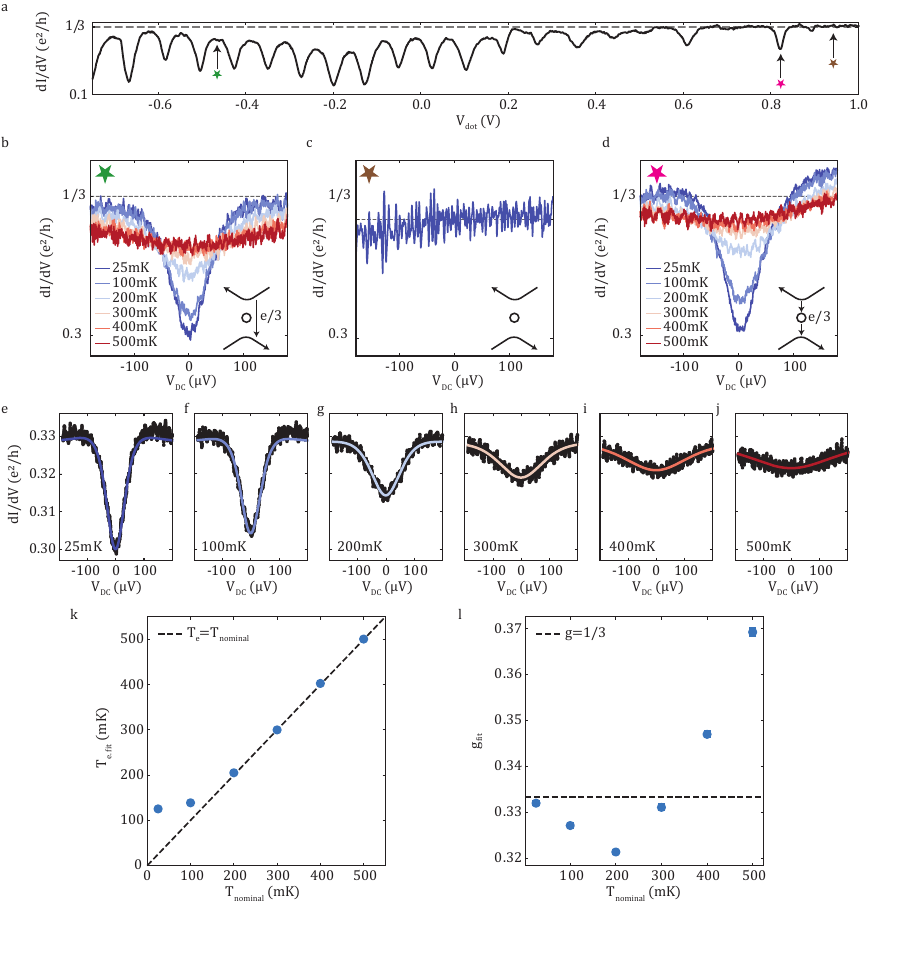} 
\caption{\textbf{Luttinger-liquid behavior of the fractional dot.}
\textbf{a,} Zero-bias differential conductance $\mathrm{d}I/\mathrm{d}V$ as a function of $V_{\rm dot}$, measured at $B=13.5~\mathrm{T}$, $V_{\rm T}=0.08~\mathrm{V}$, $V_{\rm B}=0.3~\mathrm{V}$, $V_{\rm qpc}=-0.4~\mathrm{V}$, and $V_{\rm edge}=-1.25\,\mathrm{V}$.
\textbf{b,} Bias dependence of $\mathrm{d}I/\mathrm{d}V$ at $V_{\rm dot}=-0.46~\mathrm{V}$, measured at different temperatures. At this gate voltage, the dot is in Coulomb blockade, but the QPC retains finite transmission, as illustrated in the inset.
\textbf{c,} Bias dependence of $\mathrm{d}I/\mathrm{d}V$ at $V_{\rm dot}=0.95~\mathrm{V}$. In this regime, no Coulomb resonances are observed; the QPC is fully transmitting, and the conductance is nearly independent of bias.
\textbf{d,} Bias dependence of $\mathrm{d}I/\mathrm{d}V$ at $V_{\rm dot}=0.88~\mathrm{V}$ and $V_{\rm edge}=-1.15~\mathrm{V}$, measured at different temperatures. Here, $V_{\rm dot}$ is adjusted to tune the dot to the center of a resonance (or a charge-degeneracy point) while maintaining a transmission similar to that in \textbf{b}, as shown in the inset.
\textbf{e--j,} Fits to the chiral-Luttinger-liquid model with the quasiparticle charge fixed to $e^*=e/3$. The characteristic temperature $T_0$ and background conductance $G_0$ are determined by fitting the $300~\mathrm{mK}$ data while assuming an electron temperature of $T_{\rm e}=300~\mathrm{mK}$ and are subsequently held fixed for the other temperatures.
\textbf{k,} Electron temperature $T_{\rm e}$ extracted from the fits as a function of the nominal mixing-chamber temperature $T_{\rm nominal}$. The extracted and nominal temperatures agree well at high temperatures, whereas $T_{\rm e}$ saturates at approximately $110~\mathrm{mK}$ at low temperatures, indicating imperfect electron thermalization.
\textbf{l,} Luttinger-liquid parameter $g$ extracted from the fits at different temperatures. The extracted values roughly agree with the theoretical prediction of $g=\frac{1}{3}$.}
\label{fig:Luttinger} 
\end{figure*}

In the first electron-beam lithography (EBL) step, alignment crosses were patterned and etched approximately 5~nm into the upper hBN layer. The etching was performed for 10~s using a CHF$_3$/O$_2$ gas mixture with flow rates of 40/3.5~sccm, an RF power of 160~W, and a chamber pressure of 10~mTorr.\\
Atomic-force-microscopy-based scanning microwave impedance microscopy (AFM-SMIM) was subsequently used to image the graphite top gates buried beneath the upper hBN layer. The resulting microwave-impedance images enabled the subsequent metallic gates to be aligned with the buried graphite gates, which could not be reliably achieved using topographic features alone, as shown in \ref{fig:device}f--i.

In the second EBL step, an etch mask defining the device geometry and separating the graphite top gates was patterned. A 50-nm-thick Al layer was deposited by electron-beam evaporation at a rate of 0.5~nm\,s$^{-1}$ and patterned by lift-off. The heterostructure was then etched for 4~min using CHF$_3$/O$_2$ flow rates of 40/3.5~sccm, an RF power of 160~W, and a chamber pressure of 10~mTorr. Aluminum was removed by leaving the device for an hour in AZ MIF 300 and 20 minutes in DI water.

In the third EBL step, the electrical contacts were patterned. Before metal deposition, the contact regions were exposed by 30~s of RIE. A Ti/Pd/Au metal stack with thicknesses of 5/15/130~nm was subsequently deposited and patterned by lift-off; see optical image in \ref{fig:device}j.

Finally, in the fourth EBL step, metallic finger gates were aligned with the buried graphite top gates using the AFM-SMIM images as a reference. The gates consisted of 50-nm-thick Al deposited at a rate of 0.5~nm s$^{-1}$ and were patterned by lift-off; see \ref{fig:device}k.

\subsection{Measurements}

Measurements were performed in a dry dilution refrigerator with a base temperature of approximately 20~mK. To suppress high-frequency noise and reduce the effective electron temperature, the transport lines were equipped with commercial $\pi$ filters and two stages of RC filters. The first stage consisted of two RC filters with component values of 500~$\Omega$/2.2~nF and 1.2~k$\Omega$/1.0~nF, corresponding to nominal cutoff frequencies of approximately 145 and 133~kHz, respectively. A second filtering stage, located closer to the sample, consisted of a 10~k$\Omega$/2.2~nF RC filter with a cutoff frequency of approximately 7.2~kHz.  The device was measured during two separate cool downs. During the first cool down, electronic transport measurements were performed using standard lock-in techniques. For differential-conductance measurements, an AC voltage excitation at 13.333 or 17.777~Hz was applied through a resistive voltage divider. The resulting current was measured using a Basel Precision Instruments transimpedance amplifier (SP983c), while the voltage drop across the device was simultaneously recorded using a low-noise voltage preamplifier from Basel Precision Instruments (SP1004). During the second cool down, the measurements were performed without lock-in detection. The DC current and voltage signals were generated and acquired using digital-to-analog and analog-to-digital converters (DACs and ADCs), respectively. The differential conductance was subsequently obtained numerically from the measured current--voltage characteristics using a multipoint derivative. Measurement parameters and gate configurations for all panels in the main text are listed in \ref{tab:figure1j_linecut_settings}.

\begin{table*}[t]
  \caption{\textbf{Gate configurations for main text figures}. This table contains the filling factor, gate voltages, out-of-plane magnetic field, and the measurement frequency for all the main text figures. }
  \label{tab:figure1j_linecut_settings}
  \centering
  \begin{ruledtabular}
    \footnotesize
    \setlength{\tabcolsep}{3pt}
    \begin{tabular}{llccccccccc}
      Figure
      & State
      & $B$ (T)
      & $V_{\mathrm{dot}}$ window (V)
      & $V_{\mathrm{BG}}$ (V)
      & $V_T$ (V)
      & $V_{\mathrm{qpc}}$ (V)
      & $V_{\mathrm{\rm edge}}$ (V)
      & $V_{\mathrm{AC}}$ ($\mu$V)
      & $V_{\mathrm{DC}}$ (mV)
      & Frequency \\
      \hline
      1f & $\nu=1$, decoupled & $5$ & $[-2.5,\,-0.5]$ & $0.30$ & $0.080$ & $-3.0$ & $1.5$ & N/A & $0$ & DC \\
         & $\nu=1$, coupled   & $5$ & $[-2.5,\,-0.5]$ & $0.30$ & $0.080$ & $-3.0$ & $-1.5$ & N/A & $0$ & DC \\
      \hline
      1h & \shortstack[l]{$\nu_{\mathrm{\rm bulk}}=2/5$, $\nu_{\mathrm{dot}}=1/3$\\decoupled} & $13.5$ & $[-1.5,\,0.5]$ & $0.30$ & $0.080$ & $-0.40$ & $-0.50$ & N/A & $0$ & DC \\
         & \shortstack[l]{$\nu_{\mathrm{\rm bulk}}=2/5$, $\nu_{\mathrm{dot}}=1/3$\\coupled}   & $13.5$ & $[-1.5,\,0.5]$ & $0.30$ & $0.080$ & $-0.40$ & $-1.25$ & N/A & $0$ & DC \\
      \hline
      1j & $\nu=1$, QP   & $5$    & $[2.4,\,4.4]$   & $0.30$ & $0.080$ & $-3.0$  & $-3.5$  & N/A & $1.5$ & DC \\
         & $\nu=1$, QH   & $5$    & $[-3.0,\,-1.0]$ & $0.30$ & $0.080$ & $-3.0$  & $-3.5$  & N/A & $0.09$ & DC \\
         & $\nu=1$, QH   & $8.97$ & $[0.5,\,2.5]$   & $0.0$  & $0.385$ & $-3.0$  & $-2.0$  & $68$ & $0$ & $13.77~\mathrm{Hz}$ \\
         & $\nu=2/3$, QP & $8.97$ & $[3.0,\,5.0]$   & $0.0$  & $0.305$ & $-3.0$  & $0.75$ & $6.8$ & $0$ & $17.77~\mathrm{Hz}$ \\
         & $\nu=1$, QH   & $13.4$ & $[0.0,\,2.0]$   & $0.30$ & $0.350$ & $-0.40$ & $-4.0$  & $13.6$ & $0$ & $13.77~\mathrm{Hz}$ \\
         & $\nu=2/3$, QP & $13.5$ & $[2.5,\,4.5]$   & $0.30$ & $0.220$ & $-0.40$ & $-1.25$ & N/A & $0.05$ & DC \\
         & $\nu=1/3$, QH & $13.5$ & $[-2.0,\,0.0]$  & $0.30$ & $0.080$ & $-0.40$ & $-1.25$ & N/A & $0.055$ & DC \\
      \hline
      3a & $\nu_{\mathrm{\rm bulk}}=1/3$ & $13.5$ & $[-1.03,\,3.97]$ & $0.29$ & $0.060$ & $-3.0$ & $-4.0$ & N/A & $[-1.5,\,1.5]$ & DC \\
      \hline
      4a & $\nu=1/3$, QH & $13.5$ & $[-0.9,\,0.1]$ & $0.30$ & $0.080$ & $-0.40$ & $-1.25$ & N/A & $[-1.5,\,1.5]$ & DC \\
      \hline
      4b & $\nu=2/3$, QP & $13.5$ & $[3.5,\,4.5]$ & $0.30$ & $0.220$ & $-0.40$ & $-1.25$ & N/A & $[-0.5,\,0.5]$ & DC \\
      \hline
      4c & $\nu=1/3$, QP & $13.5$ & $[3.4,\,4.4]$ & $0.30$ & $0.050$ & $-0.40$ & $-4.0$ & N/A & $[-1.5,\,1.5]$ & DC \\
      \hline
      4d & $\nu=2/3$, QH & $13.5$ & $[-1.0,\,0.0]$ & $0.30$ & $0.270$ & $-0.40$ & $-4.0$ & N/A & $[-0.5,\,0.5]$ & DC \\
      \hline
      4e & $\nu=1/3$, QP & $13.5$ & $[3.8,\,4.3]$ & $0.29$ & $0.050$ & $-3.0$ & $-4.0$ & N/A & $[-1.5,\,1.5]$ & DC \\
      \hline
      4f & $\nu=1/3$, QP & $13.5$ & $[3.785,\,4.5]$ & $0.29$ & $[0.040,\,0.050]$ & $-3.0$ & $-4.0$ & N/A & $0.055$ & DC \\
      \hline
      4g  & $\nu=1/3$, QP & $13.5$ & $[4.0,\,4.5]$ & $0.29$ & $0.040$ & $-3.0$ & $-4.0$ & N/A & $[-1.5,\,1.5]$ & DC \\
    \end{tabular}
  \end{ruledtabular}
\end{table*}

\section{Theoretical calculations}

\subsection{Device model}

To obtain a realistic external electrostatic potential, we solved the electrostatics of the device with our own axisymmetric finite-difference Poisson solver. The simulated geometry reproduced the gate structure of the measured quantum dot, assuming a hole radius of 45~nm, extracted from a scanning electron microscope micrograph. The anisotropic dielectric response of hBN was modeled using an out-of-plane relative permittivity of $\varepsilon_{\perp}=3.0$ and an in-plane relative permittivity of $\varepsilon_{\parallel}=6.6$. The top and bottom hBN layers were assigned thicknesses of 32~nm and 50~nm, respectively, based on atomic force microscopy measurements of the flakes used in the device. The metallic gates were positioned 32~nm above the top graphite layer and separated from it by the top hBN layer. The graphene monolayer was treated as an electrically insulating plane, and the electrostatic potential was evaluated at its vertical position. In the same geometry, the solver reproduces finite-element simulations with the Electrostatics Module of COMSOL Multiphysics to $0.05\%$. 

\begin{figure*}[t]
    \centering
    \includegraphics[width=\textwidth]{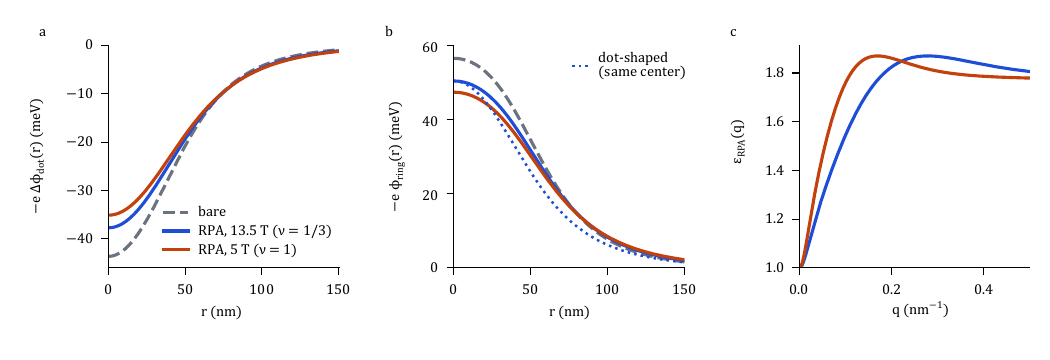}
    \caption{\textbf{Device potentials and screening.} \textbf{a}, Potential energy of an electron due to the dot gate, $-e\Delta\phi_{\rm{dot}}(r) = -e(\phi_{\rm{dot}}^{1\rm{V}}(r)-\phi_{\rm{dot}}^{0\rm{V}}(r))$, as a function of distance from the dot center $r$. The residual potential at $0~\rm{V}$ is removed. RPA is applied at the magnetic fields of interest. \textbf{b}, Potential energy of an electron due to the edge ring from a voltage step $\Delta V=-2.02~\rm{V}$ on the edge face of the top gate hole. The bare and RPA-screened potentials are included as well as a dot-shaped potential with the same center value at $B=13.5~\rm{T}$ to emphasize their distinct shapes. \textbf{c}, Dielectric function $\varepsilon_{\mathrm{RPA}}(q) = 1 - V_0(q)\,\Pi(q)$, where $V_0$ is the dual-gated Coulomb kernel and $\Pi$ the static inter-Landau-level polarization of monolayer graphene. Colors as in \textbf{a, b}. 
    }
    \label{fig:dot_potential}
\end{figure*}

\begin{figure*}
\centering
\includegraphics[]{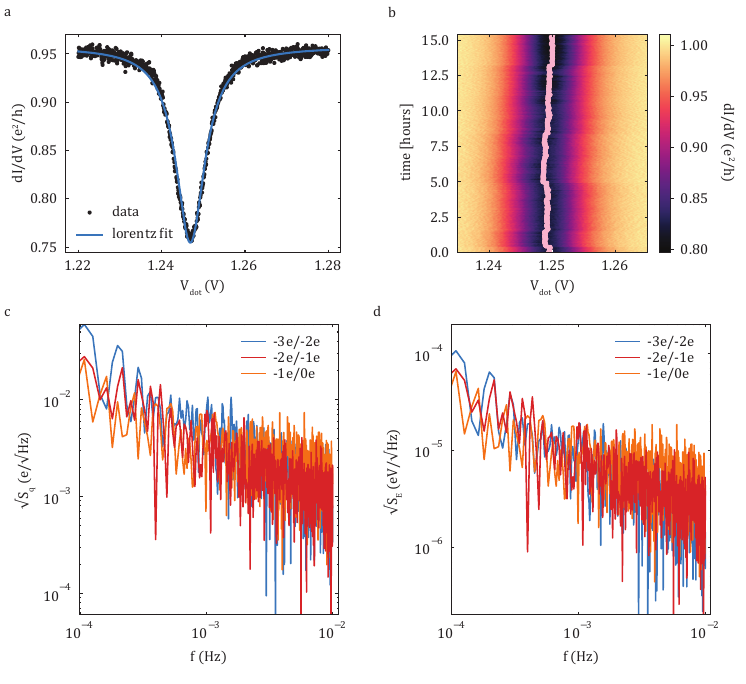}
\caption{\textbf{Charge-noise characterization of the integer dot.}
\textbf{a,} Zero-bias conductance measured across the $N=-3\rightarrow-2$ charge transition at $B=5~\mathrm{T}$, with the other gate voltages set to $V_{\rm B}=0.3~\mathrm{V}$, $V_{\rm T}=0.066~\mathrm{V}$, $V_{\rm qpc}=-3~\mathrm{V}$, and $V_{\rm edge}=-5~\mathrm{V}$. The blue curve is a Lorentzian fit to the Coulomb resonance.
\textbf{b,} Time evolution of the charge-transition position, measured continuously for 15.5 hours. The Coulomb-resonance position, indicated by the pink circles, is extracted from a fit to each trace. The small drift and fluctuations demonstrate the excellent long-term charge stability of the integer dot.
\textbf{c} [\textbf{d}], Low-frequency charge-noise spectral density for the first three quasihole transitions of the integer dot at $B=5~\mathrm{T}$, expressed in units of $e/\sqrt{\mathrm{Hz}}$ [$\mathrm{eV}/\sqrt{\mathrm{Hz}}$], respectively. The spectra were measured across the $N=-3\rightarrow-2$ transition at $V_{\rm qpc}=-3~\mathrm{V}$ and $V_{\rm edge}=-5~\mathrm{V}$; the $N=-2\rightarrow-1$ transition at $V_{\rm qpc}=-3~\mathrm{V}$ and $V_{\rm edge}=-5~\mathrm{V}$; and the $N=-1\rightarrow0$ transition at $V_{\rm qpc}=-3~\mathrm{V}$ and $V_{\rm edge}=-1.5~\mathrm{V}$. The results are summarized in \ref{tab:charge_noise}.}
\label{fig:extended_noise} 
\end{figure*}

\begin{figure*}
\centering
\includegraphics[]{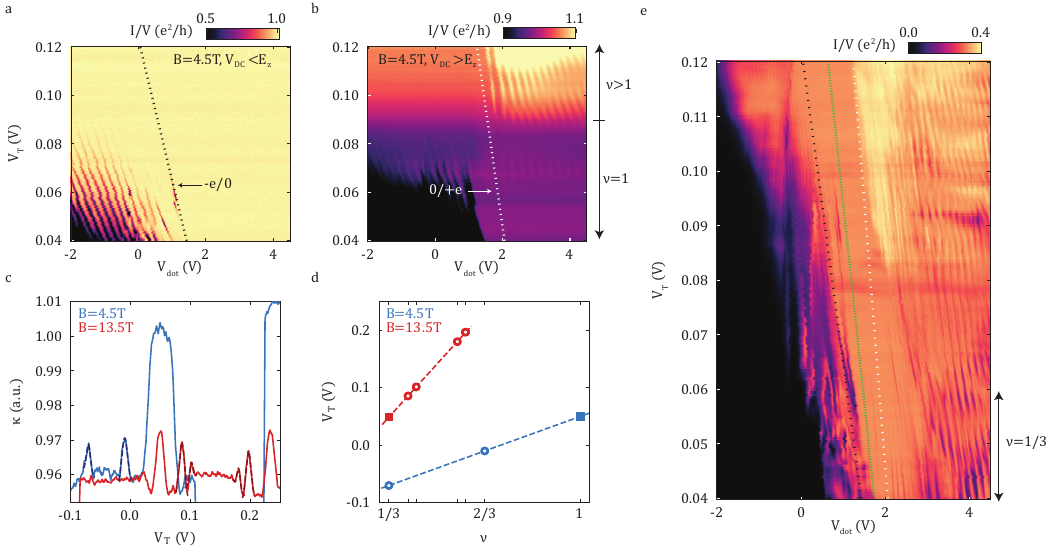} 
\caption{\textbf{Identification of the single-anyon regime.} \textbf{a} [\textbf{b}]\textbf{,} $I/V$ as a function of $V_{\mathrm{T}}$ and $V_{\mathrm{dot}}$ at $B=4.5$~T, measured using the same gate configuration as in \ref{fig:fraction}, with $V_{\mathrm{DC}}=50~\mu\mathrm{V}$ [$V_{\mathrm{DC}}=1.8$~mV]. The dashed lines identify the addition of the first quasihole in \textbf{a} and the first quasiparticle in \textbf{b}. The quasiparticle resonances are not visible at low bias, hence \textbf{b} is taken to identify the quasiparticle resonances. For $\nu>1$, quasiparticle addition produces a conductance peak rather than a dip, suggesting the presence of an additional inner edge mode that enables forward scattering. \textbf{c,} Penetration-field capacitance, $\kappa$, as a function of $V_{\mathrm{T}}$, measured using the same gate-voltage configuration as in \ref{fig:fraction}, at magnetic fields of 4.5 and 13.5~T. Gaussian fits were used to identify the $\nu=1$ and $2/3$ gaps at 4.5~T and the $\nu=2/5$, $3/7$, $4/7$, $3/5$, and $2/3$ gaps at 13.5~T. The fits are shown as dashed lines over the experimental data. The uncertainty in each fitted position was taken to be 10\% of the corresponding full width at half maximum. \textbf{d,} Filling-factor dependence of the fitted positions in $V_{\mathrm{T}}$ (open circles). Solid squares indicate the independently extracted positions of $\nu=1$ at 4.5~T and $\nu=1/3$ at 13.5~T. We find $\nu=1/3$ at $V_{\mathrm{T}}=49\pm1$~mV and $\nu=1$ at $V_{\mathrm{T}}=50\pm2.5$~mV. Their agreement constrains the uncertainty in the charge-neutrality reference to be smaller than one fractional period in $V_{\mathrm{dot}}$. \textbf{e,} Same measurement as in \textbf{a}, acquired at $B=13.5$~T. The green dashed line indicates the expected position of the zero-quasiparticle regime extrapolated from the $B=4.5 \mathrm{T}, \nu=1$ data shown in \textbf{a} and \textbf{b}. This analysis enables identification of the first quasiparticle and quasihole additions in the anyon regime.}
\label{fig:zeroqp_identification} 
\end{figure*}

The gate geometry alone does not reproduce the measured dot voltages of the plateaus. For instance, the $\nu=1/3$ plateau in the zero excitation sector is centered at the high voltage $V_{\rm{dot}}\approx 1.44~\rm{V}$. We attribute this inconsistency to the presence of electric dipoles at the etched edge of the top gate hole and model them as a work function difference on the inner edge of the top gate trenches.  We implement this by keeping the inner edge face of the hole at potential $V_T+\Delta V$ while the rest of the top gate is held at $V_T$. 
We compute the step's contribution to the potential in the graphene plane using our solver. We use a top gate thickness of $0.5~\rm{nm}$ and refine the solver's mesh to $0.001~\rm{nm}$ at the edge where this response is concentrated. We fix $\Delta V=-2.02~\rm{V}$ once at $\nu=1/3$ so that the model reproduces the measured position of the plateau and use the same step at $\nu=1$. It repels electrons by about $50~\rm{meV}$ at the dot center, which mostly shifts the dot-gate axis uniformly. The remainder is a barrier near the rim of the hole that the dot gate cannot cancel. The resulting potential landscape was subsequently used as an input for the theoretical simulations (see \ref{fig:dot_potential}a,b). 

The dual-gated planar interaction can be written
\begin{equation}
V_0(q) = \frac{2\pi\ell_B E_{\rm{Coul}}}{q}\,
\frac{2\sinh(\beta q d_b)\,\sinh(\beta q d_t)}{\sinh(\beta q\,(d_b+d_t))}
\label{eq:V0}
\end{equation}
where $\varepsilon = \sqrt{\varepsilon_\parallel\varepsilon_\perp}$ is the effective permittivity and $\beta = \sqrt{\varepsilon_\parallel/\varepsilon_\perp}$ is the anisotropy factor, and $E_{\rm{Coul}}=e^2/(4\pi\varepsilon_0\varepsilon \ell_B)$ is the Coulomb energy scale. The magnetic length is $\ell_B = 6.98$\,nm at $B = 13.5$\,T and $11.47$\,nm at $B = 5$\,T, giving $E_{\mathrm{Coul}} = 46.36$\,meV and $28.21$\,meV, respectively. We denote by $d_t$ and $d_b$ the distances from the graphene monolayer to the top and bottom gates, respectively. We include screening by virtual inter-Landau-level transitions within the random phase approximation (RPA)
\begin{equation}
V(q) = \frac{V_0(q)}{\varepsilon_{\mathrm{RPA}}(q)},
\qquad
\varepsilon_{\mathrm{RPA}}(q) = 1 - V_0(q)\,\Pi(q),
\label{eq:VRPA}
\end{equation}
where $\Pi(q)$ is the static polarization function of monolayer graphene in a magnetic field (see \ref{fig:dot_potential}c). The same dielectric function screens the dot potential. For calculations on the sphere, we use this interaction and not the analogue sphere interaction. They agree in the thermodynamic limit, but using the planar interaction makes extrapolation simpler. The real-space interaction is obtained by the Hankel transform
\begin{equation}
v(r) = \int_0^\infty \frac{q\,dq}{2\pi}\; V(q)\, J_0(qr),
\label{eq:vr}
\end{equation}
where $J_0$ is the zeroth-order Bessel function of the first kind.

\subsection{DMRG on the Haldane sphere}

We model the quantum dot with the density matrix renormalization group (DMRG) in Haldane's spherical geometry~\cite{haldane_fractional_1983, feiguin_density_2008}. In this setup, $N_e$ electrons are confined to the surface of a sphere of radius $R=\sqrt{N_\phi/2}\ell_B$. The sphere is threaded by $N_\phi$ flux quanta generated by a magnetic monopole at its center. We work in the lowest Landau level, which consists of $N_\phi+1$ orbitals labeled by the $z$ component of angular momentum, $m = -N_\phi/2, \dots, N_\phi/2$, each localized around a polar angle that increases monotonically with $-m$. We place the dot at the north pole. Since the dot potential is axially symmetric, it conserves $L_z$, enabling us to split the Hilbert space into decoupled $L_z$ sectors. The one-body energies follow from the screened dot potential, and the screened Coulomb interaction is evaluated at the chord distance between electrons. We map the orbitals onto a one-dimensional chain ordered by $m$ and run finite DMRG within physically motivated $L_z$ sectors. For every charge $q$, we scan $L_z$ and take the lowest energy over all sectors as the ground state. 

We present the formalism for the Laughlin state at $\nu=1/3$~\cite{laughlin_anomalous_1983}. It applies analogously to the integer state at $\nu=1$ with excitations of charge $e^*=e$. At $\nu=1$, added electrons occupy the opposite spin level and pay the Zeeman energy, and therefore we include the spin degree of freedom in the DMRG. The Laughlin liquid at $\nu=1/3$ is the unique ground state at $N_\phi=3(N_e-1)$ with total angular momentum $L=0$~\cite{haldane_fractional_1983}. Removing (adding) $q$ flux quanta at fixed $N_e$ creates $q$ quasiparticles (quasiholes) of charge $e^*=e/3$ ($-e/3$) where $e<0$ is the electron charge~\cite{laughlin_anomalous_1983, haldane_fractional_1983}. At $\nu=1/3$, we treat the electrons as fully spin- and valley-polarized. For the DMRG, we use $N_e=8$--$18$ electrons and up to $q=6$ quasiparticles. For $\nu=1$ we use sizes $N_e=20$--$50$. We increase the bond dimension in steps up to a maximum of $\chi=1600$. Sweeps continue until the energy changes by less than $2\times 10^{-6}E_{\rm{Coul}}$ over five consecutive sweeps. Increasing the bond dimension to $\chi=3200$ changes the plateau widths by less than $0.2~\rm{mV}$. We then extrapolate to the thermodynamic limit $N_e\rightarrow\infty$, which we now discuss. 

Comparing sectors consists of a few steps with the goal of isolating the energy of the excess charge. First, the sphere has no neutralizing background, so the ground state energy $E_q$ contains the extensive Hartree energy of the entire uniform liquid. We therefore add a rigid, uniform background with charge $eN_{\rm{ref}}=e(N_e-q/3)$ to compensate. The corrected energy can then be written as
\begin{equation}
\tilde{E}_q = E_q - N_{\rm{ref}}\overline{U} - \frac{\overline{v}}{2}\left(2N_e N_{\rm{ref}} - N_{\rm{ref}}^2\right)
\label{eq:corrected_gs_energy}
\end{equation}
where $\overline{U}$ is the one-body potential averaged over the sphere orbitals and $\overline{v}$ is the interaction averaged over the sphere. The second term is the background's energy in the external potential, and the third is its Hartree interaction with the electrons together with its self-energy. Since $2N_eN_{\rm{ref}} - N_{\rm{ref}}^2 = N_e^2 - (q/3)^2$, the subtraction removes the charging energy of the liquid but keeps that of the excess charge. 

Each state is computed at two or more dot voltages $V_c$ that bracket its transitions. Since the one-body energies $U_m$ are linear in $V_{\mathrm{dot}}$, we evaluate the energy at other voltages to first order,
\begin{equation}
\tilde E(V_{\mathrm{dot}}) = \tilde E(V_c) + (V_{\mathrm{dot}} - V_c)\sum_m \Big(n_m - \frac{N_{\mathrm{ref}}}{N_\phi + 1}\Big)\frac{\partial U_m}{\partial V_{\mathrm{dot}}},
\label{eq:Elin}
\end{equation}
where $n_m$ are the orbital occupations at $V_c$. The term $N_{\mathrm{ref}}/(N_\phi+1)$ is the occupation of the uniform background, so the dot gate couples only to the excess charge. We take the lowest estimate from the different $V_c$. Relaxation of the states changes these energies by less than $3\times10^{-4}\,E_{\mathrm{Coul}}$ over 300~mV, so this interpolation is reliable for competing energies much larger than this amount. 

The dot exchanges charge with the surrounding system which acts as a reservoir with electrochemical potential $\mu$ per charge $e^*=\nu e$. The ground state minimizes the free energy
\begin{equation}
G_q(V_{\rm{dot}}) = \min_{L_z}\tilde E_q(V_{\rm{dot}}) - q\,\mu ,
\label{eq:Gq}
\end{equation}
where $q$ counts the excess charge in units of $e^*$. In the absence of a dot potential, the incompressible state is the ground state for $\tilde E^{(0)}_0 - \tilde E^{(0)}_{-1} < \mu < \tilde E^{(0)}_{+1} - \tilde E^{(0)}_0$, whose center is $\mu^* = \frac12\big(\tilde E^{(0)}_{+1} - \tilde E^{(0)}_{-1}\big)$. In experiment, the bulk is not exactly tuned to the center of the incompressible plateau, and so we set $\mu = \mu^*+\delta\mu$ with $\delta\mu$ motivated from the top gate voltage relative to the center of the plateau. We use $\delta\mu=5.15~\rm{meV}$ per electron ($\delta\mu=1.72~\rm{meV}$ per $e/3$) at $\nu=1/3$ and $\delta\mu=6~\rm{meV}$ at $\nu=1$. At the latter filling, $\mu^*$ includes the Zeeman energy. At the transition voltages defined by $G_q(D_q)=G_{q-1}(D_q)$, the dot is tuned to a charge degeneracy between $q$ and $q-1$. The plateau width of charge sector $q$ is then defined as $W_q=D_{q+1}-D_q$. 

Since the lowest-energy state can change with system size, we extrapolate each angular momentum sector separately. Defining each energy with respect to that of the Laughlin state of the same size (which removes the extensive correlation energy), we fit
\begin{equation}
\tilde E_{q,L_z} - \tilde E_{0,0} = \epsilon^{\infty}_{q,L_z} + \frac{c_1}{N_e} + \dots
\label{eq:extrap}
\end{equation}
pointwise in $V_{\rm{dot}}$. We extrapolate $\mu^*$ in the same way. We then minimize the extrapolated energy $\epsilon^{\infty}_{q,L_z}$ over $L_z$ and solve for $D_q$. 

\ref{fig:cf_filling} shows the composite-fermion configurations of the quasiparticle ground states, identified by their overlaps with projected Jain wavefunctions~\cite{jain_composite-fermion_1989,jain_composite_2007}. The first three quasiparticles fill $\Lambda_1$ from the dot center outward. The fourth and fifth enter $\Lambda_2$ near the center, and the sixth returns to $\Lambda_1$. A higher $\Lambda$ level costs energy but keeps the charge closer to the dot center. This competition is the mechanism behind the irregular filling sequence we find. We denote orbital $k$ of $\Lambda$ level $i$ as $\Lambda_i(k)$, and order orbitals using distance from dot center~\cite{jain_composite_2007}. The pairs $\Lambda_1(2)/\Lambda_2(0)$ and $\Lambda_2(1)/\Lambda_1(3)$ are nearly degenerate, so the plateaus that separate them at $e$ and $5e/3$ are short.

\subsection{Molecular anyons in a quantum dot}

In the absence of an external potential, $n$ like-charged anyons bind when
\begin{equation}
E_{\mathrm{bind}}^{(n)} = nE_1 - E_n > 0 ,
\end{equation}
where $E_n$ is the energy of the liquid carrying $n$ anyons, measured from the ground state~\cite{wang_anyon_2026}. If the anyons repel, they separate indefinitely and $E_n = nE_1$. 
With this sign convention, $E_{\mathrm{bind}}^{(n)} > 0$ means the cluster has lower energy than $n$ isolated anyons. 
We focus on pairs, $n = 2$. 
In the presence of a dot, the anyons are confined and so there is no notion of infinite separation. 
Pairing is better understood in terms of energy convexity with respect to the number of trapped anyons, $q$. T
he addition energy can be written as the second difference
\begin{equation}
    E_{\rm{add}}(q) = G_{q+1}-2G_q+G_{q-1}, 
\end{equation}
where $G_q$ is the free energy (Eq.~\ref{eq:Gq}) of the system with $q$ trapped anyons. 
If $E_{\rm{add}}(q)>0$, then the state with $q$ anyons is the ground state over a finite range of $V_{\rm{dot}}$. 
If $E_{\rm{add}}(q)<0$, then that state is never the ground state and the charge jumps directly from $q-1$ to $q+1$. 
Pairing requires the inter-anyon attraction to overcome the confinement and charging cost. 
Short plateaus or an even/odd pattern in the widths do not imply binding (as they may due to attraction or orbital effects). In addition, binding in the dot does not imply binding in a uniform system. 
In our calculations $E_{\mathrm{add}}(q) > 0$ for the few quasiparticles we compute, so the charge always enters in single steps of $e^*$. Intrinsic binding in a uniform system may nonetheless be related to the mechanism that would enable binding in the dot, and so we discuss it below. 

Intrinsic binding energies are small (of order $10^{-3}\,E_{\mathrm{Coul}}$) and require large systems. On the sphere, the pair binding at $\nu = 2/5$ is still far from converged at $N_e \le 20$. We therefore compute bulk binding energies with segment DMRG on an infinite cylinder without the dot~\cite{zaletel_topological_2013,wang_anyon_2026}. In contrast, the plateau widths are dominated by much larger addition energies and converge well on the sphere. The dot potential fixes the separation of the anyons and raises the local filling as charge accumulates, so the effective pair interaction can differ from its bulk value. 
The bulk binding energies therefore indicate where pairing is favorable, not the precise conditions under which pairs form in the dot. 

The liquid is placed on an infinite cylinder of circumference $L_y$ in the lowest Landau level and in the Landau gauge where orbital $k$ is centered at $x_k = 2\pi k\ell_B^2/L_y$. We use the gated, RPA-screened interaction evaluated at the momenta allowed by the cylinder. The interaction is represented as a matrix-product operator (MPO) compressed to a tolerance that changes the binding energies by less than $10^{-6}\,E_{\mathrm{Coul}}$. The ground state is obtained with infinite DMRG in fixed charge and momentum sectors, for $L_y = 16$--$28\,\ell_B$ and bond dimensions up to $\chi = 8192$~\cite{zaletel_topological_2013}.

An excitation is created by joining the ground state to a copy of itself translated by $s$ orbitals~\cite{wang_anyon_2026}. Since each orbital holds charge $\nu e$, the mismatch leaves a charge $s\nu e$ (modulo $e$) at the boundary. The charge sector of the segment determines which of these charges is realized. At $\nu=p/(2p+1)$, a translation by one orbital $(s=1)$ gives the elementary charge only for $p=1$. Therefore, to model the elementary charges, $\pm e^*$, we use $\abs{s}=1$ at $\nu=1/3$ and $\abs{s}=2$ at $\nu=2/5$. For the pair charges, $\pm 2e^*$, we use $\abs{s}=1$ at $\nu=1/3$ and $2/5$. The sign of $s$ is fixed by the charge. At $\nu=1/3$, the pair and single charges use translations of opposite sign, so that $2e/3=e-e/3$ belongs to the class of $-e/3$. We optimize a finite segment around the boundary, $34$--$50\,\ell_B$ long, embedded between the semi-infinite ground states, in a chosen charge and momentum sector. Changing the segment length alters the energies by less than $10^{-6}\,E_{\mathrm{Coul}}$. On a cylinder, a bound pair can arrange itself in several configurations with different momenta $K$, so we scan $K = -2, \dots, +2$ and take the lowest energy. Each charge also interacts with its periodic images around the circumference, so we subtract this energy using the measured charge profiles and the planar interaction and extrapolate in $\chi$.

We compute the pair binding energies of quasiparticles and quasiholes at $\nu = 1/3$ and at the Jain state $\nu = 2/5$~\cite{jain_composite-fermion_1989} (see \ref{tab:binding}). At $\nu = 1/3$, neither quasiparticles nor quasiholes bind: the pair energy is positive at every separation we sample (7.6--14$\,\ell_B$). At $\nu = 2/5$ only quasiparticles bind. In CF language, a quasiparticle at $\nu = p/(2p+1)$ is a CF in $\Lambda_p$ and a quasihole is a vacancy in $\Lambda_{p-1}$~\cite{jain_composite_2007}. Binding thus appears for CFs in $\Lambda_2$, the level occupied by the fourth and fifth quasiparticles in the dot at $\nu = 1/3$ (Fig.~\ref{fig:cf_filling}). We have studied pairs only; larger clusters may also bind. 

\begin{table}[h]
\centering
\begin{tabular}{c c c c}
\toprule
$\nu$ & $e^*$ & quasiparticle pair & quasihole pair \\
\midrule
$1/3$ & $e/3$ & unbound & unbound \\
$2/5$ & $e/5$ & $88\pm5$ & unbound \\
\bottomrule
\end{tabular}
\caption{Bulk pair binding energies $E_{\mathrm{bind}}^{(2)}$ in $\mu$eV ($1\,\mu\mathrm{eV} = 2.16\times10^{-5}\,E_{\mathrm{Coul}}$ at $B = 13.5$\,T) calculated using segment DMRG on an infinite cylinder. The uncertainty reflects the extrapolation in $\chi$.}
\label{tab:binding}
\end{table}


\clearpage
\bibliographystyle{science}
\bibliography{references}

\begin{table*}[t]
\centering
\caption{\textbf{Charge-noise summary at $B=5~\mathrm{T}$.}
Energy-noise amplitude $\sqrt{S_E}$ and equivalent charge-noise amplitude
$\sqrt{S_q}$ evaluated at selected frequencies for three consecutive
charge transitions. Data are shown in \ref{fig:extended_noise}.}
\label{tab:charge_noise}
\setlength{\tabcolsep}{0pt}
\renewcommand{\arraystretch}{1.5}
\begin{tabular*}{\textwidth}{
    @{\extracolsep{\fill}}
    c ccc ccc
}
\hline
& \multicolumn{3}{c}{$\sqrt{S_E}\;(\mathrm{eV}/\sqrt{\mathrm{Hz}})$}
& \multicolumn{3}{c}{$\sqrt{S_q}\;(e/\sqrt{\mathrm{Hz}})$} \\
\cline{2-4}\cline{5-7}

Charge transition
& $10^{-4}\,\mathrm{Hz}$
& $10^{-3}\,\mathrm{Hz}$
& $10^{-2}\,\mathrm{Hz}$
& $10^{-4}\,\mathrm{Hz}$
& $10^{-3}\,\mathrm{Hz}$
& $10^{-2}\,\mathrm{Hz}$ \\
\hline

$-3\rightarrow-2$
& $4.2\times10^{-5}$
& $8.5\times10^{-6}$
& $1.7\times10^{-6}$
& $2.4\times10^{-2}$
& $4.8\times10^{-3}$
& $9.6\times10^{-4}$ \\

$-2\rightarrow-1$
& $3.6\times10^{-5}$
& $8.7\times10^{-6}$
& $2.1\times10^{-6}$
& $1.5\times10^{-2}$
& $3.5\times10^{-3}$
& $8.5\times10^{-4}$ \\

$-1\rightarrow0$
& $2.6\times10^{-5}$
& $1.0\times10^{-5}$
& $3.9\times10^{-6}$
& $1.0\times10^{-2}$
& $4.0\times10^{-3}$
& $1.6\times10^{-3}$ \\
\hline
\end{tabular*}
\end{table*}

\begin{figure*}[p]
\centering
\includegraphics[
    width=\textwidth,
    height=0.95\textheight,
    keepaspectratio
]{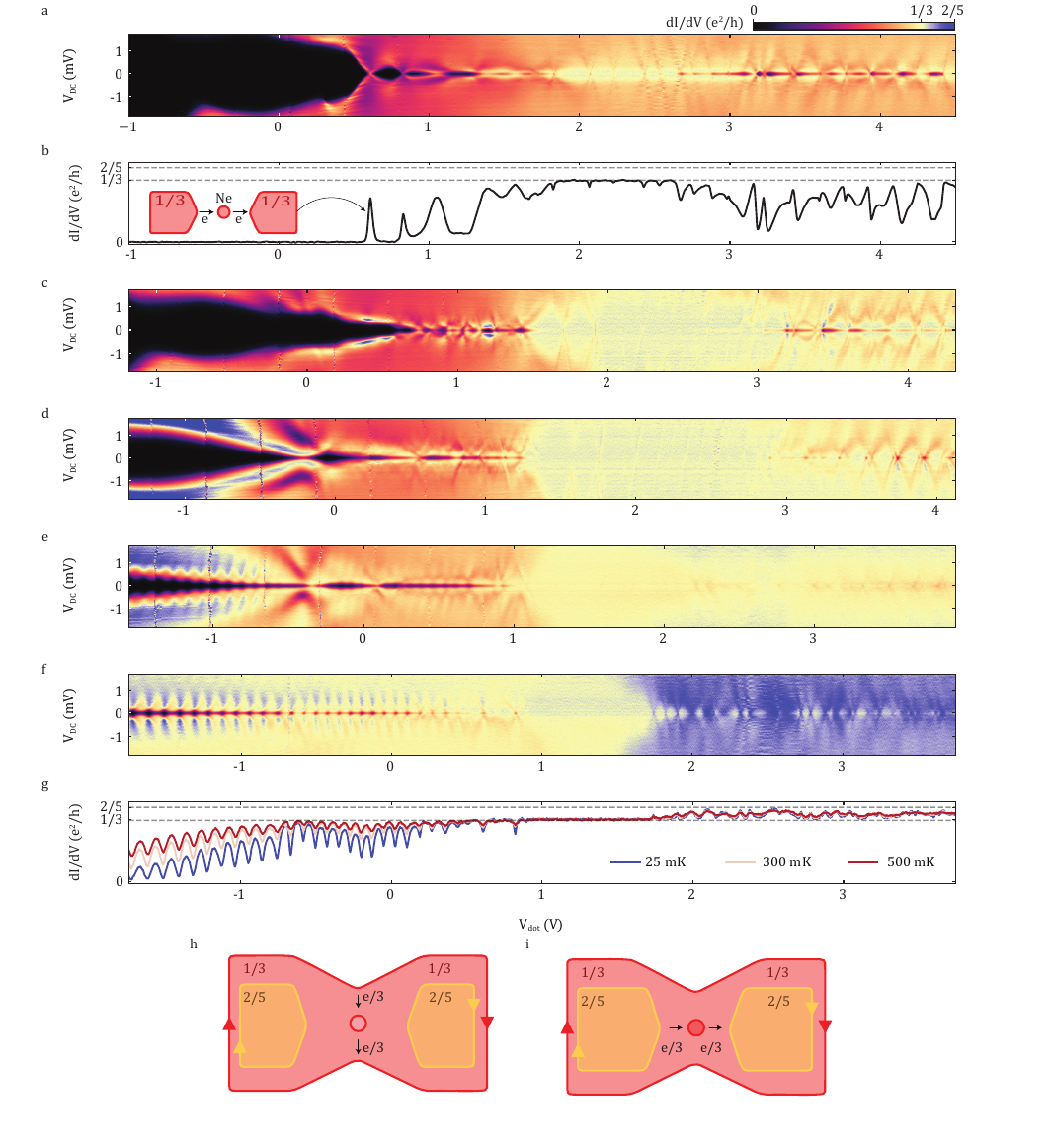}
\end{figure*}

\begin{figure*}[p]
\caption{\textbf{Virtual-gate dependence of the fractional dot.}
\textbf{a,} Differential conductance $\mathrm{d}I/\mathrm{d}V$ as a function of $V_{\rm dot}$ and DC bias $V_{\rm DC}$, measured with the bulk at $\nu_{\rm bulk}=1/3$. The device is tuned to $V_{\rm edge}=-4~\mathrm{V}$ and $V_{\rm qpc}=-3.0~\mathrm{V}$ at $B=13.5~\mathrm{T}$.
\textbf{b,} Zero-bias linecut taken from \textbf{a}. The QPC starts to pinch off around $V_{\rm dot}=1.8\mathrm{V}$ and is fully pinched off at $0.5\mathrm{V}$, as the transmission vanishes. Before the QPC fully pinches off, we see peaks in conductance, which suggests forward scattering of electrons across the QPC. As we increase the $V_{\rm dot}$ to more positive values, we see dips, which correspond to the backscattering of $e/3$ quasiparticles. This interpretation is also consistent with the Luttinger liquid suppression at low bias and the shorter spacings between resonances. As discussed in the main text, we also observe the even-odd effect at more positive $V_{\rm dot}$ values, which indicates the formation of molecules from the quasiparticles.
\textbf{c--f,} Evolution of the bias spectroscopy as the gate voltages are varied in equally spaced steps between the configurations shown in \textbf{a} and \textbf{f}.
\textbf{f,} Differential conductance measured at $V_{\rm T}=0.08~\mathrm{V}$, $V_{\rm B}=0.3~\mathrm{V}$, $V_{\rm qpc}=-0.4~\mathrm{V}$, and $V_{\rm edge}=-1.25~\mathrm{V}$, for which the bulk is tuned to $\nu_{\rm bulk}=2/5$. In each panel, $V_{\rm dot}$ is adjusted using a virtual-gate correction to track the same sequence of quasiparticle transitions. Effectively, the density range in the dot is kept fixed while other gates are tuned to adjust the coupling to the dot, hence making the QPC more transmissive. \textbf{g,} Zero-bias line cuts of \textbf{f} measured at different temperatures. On the quasihole side, in comparison to the low transmission accompanied with forward scattering of whole electrons in the first panel, we see equally spaced dips, which is interpreted as the backscattering of $e/3$ quasiparticles. On the quasiparticle side, we observe peaks in conductance which correspond to the forward scattering of $e/3$ quasiparticles. \textbf{h-i,} The two schematics further describe the two scenarios described above. The left schematic (\textbf{h}) illustrates backscattering of $e/3$ quasiparticles when the bulk is at $\nu_{\rm bulk}=2/5$ and the dot remains at $\nu_{\rm dot}=1/3$. The right schematic (\textbf{i}) illustrates forward scattering from the inner $1/15$ edge channel through a dot hosting charge-$e/3$ quasiparticles. Conductance peaks with $\mathrm{d}I/\mathrm{d}V< 2/5 ~(e^2/h)$ are observed.}
\label{fig:virtual}
\end{figure*}

\begin{figure*}[t]
    \centering
    \includegraphics[width=\textwidth]{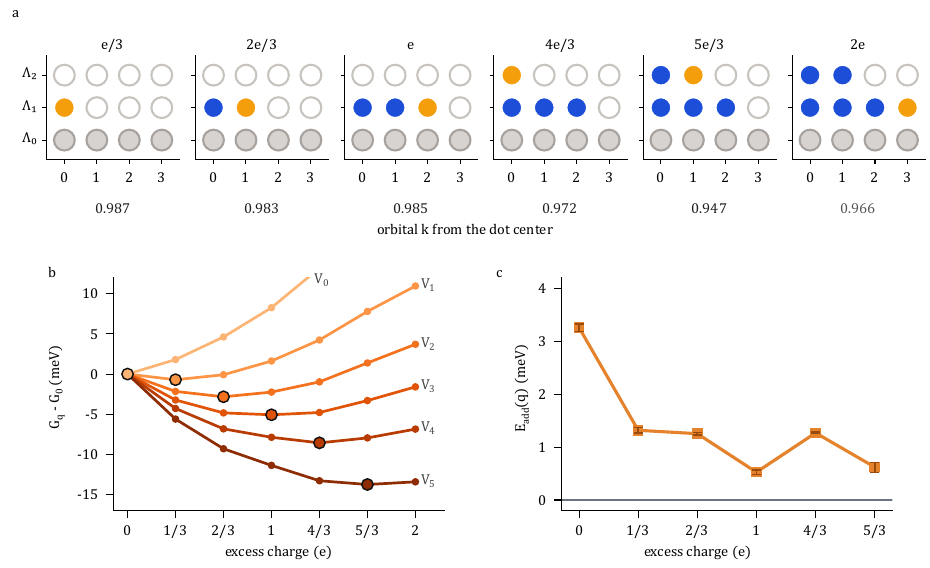}
    \caption{\textbf{Composite-fermion structure and quasiparticle charging in the dot. }\textbf{a}, $\Lambda$-level filling order as quasiparticles are added to the dot. Rows are $\Lambda$ levels and columns are the orbital index $k$, counted from the dot center. The filled $\Lambda_0$ corresponds to $\nu=1/3$. Blue denotes occupied CF orbitals and orange denotes those newly occupied relative to the previous charge. Titles give the dot charge and numbers below each panel are the overlaps $|\langle\Psi_{\mathrm{DMRG}}|\Psi_{\mathrm{CF}}\rangle|$ between the DMRG ground state of that charge sector and the projected Jain wavefunction of the configuration shown, at $N_e = 8$ for $e/3$ to $4e/3$ and $N_e = 10$ for $5e/3$ and $2e$.
    \textbf{b,c}, Convexity of the free energy at $\nu=1/3$, computed with DMRG on the Haldane sphere and extrapolated to the thermodynamic limit $N_e\rightarrow\infty$. \textbf{b}, Free energy $G_q - G_0$ (Eq.~\ref{eq:Gq}) of the system with $q$ quasiparticles in the dot, at gate voltages $V_k$ fixed at the center of plateau $k=0,\dots,5$. Each curve is convex with its minimum (outlined) at $q=k$, so every charge state is the ground state over a finite range of $V_{\rm dot}$. \textbf{c}, Addition energy $E_{\rm add}(q) = G_{q+1}-2G_q+G_{q-1}$ at the center of plateau $q$, with error bars from the extrapolation (smaller than the symbols where not visible). Since $E_{\rm add}(q)>0$ for all $q$, the charge always enters in single steps of $e^*$.}
    \label{fig:cf_filling}
\end{figure*}

\begin{figure*}
\centering
\includegraphics[width=5.7in]{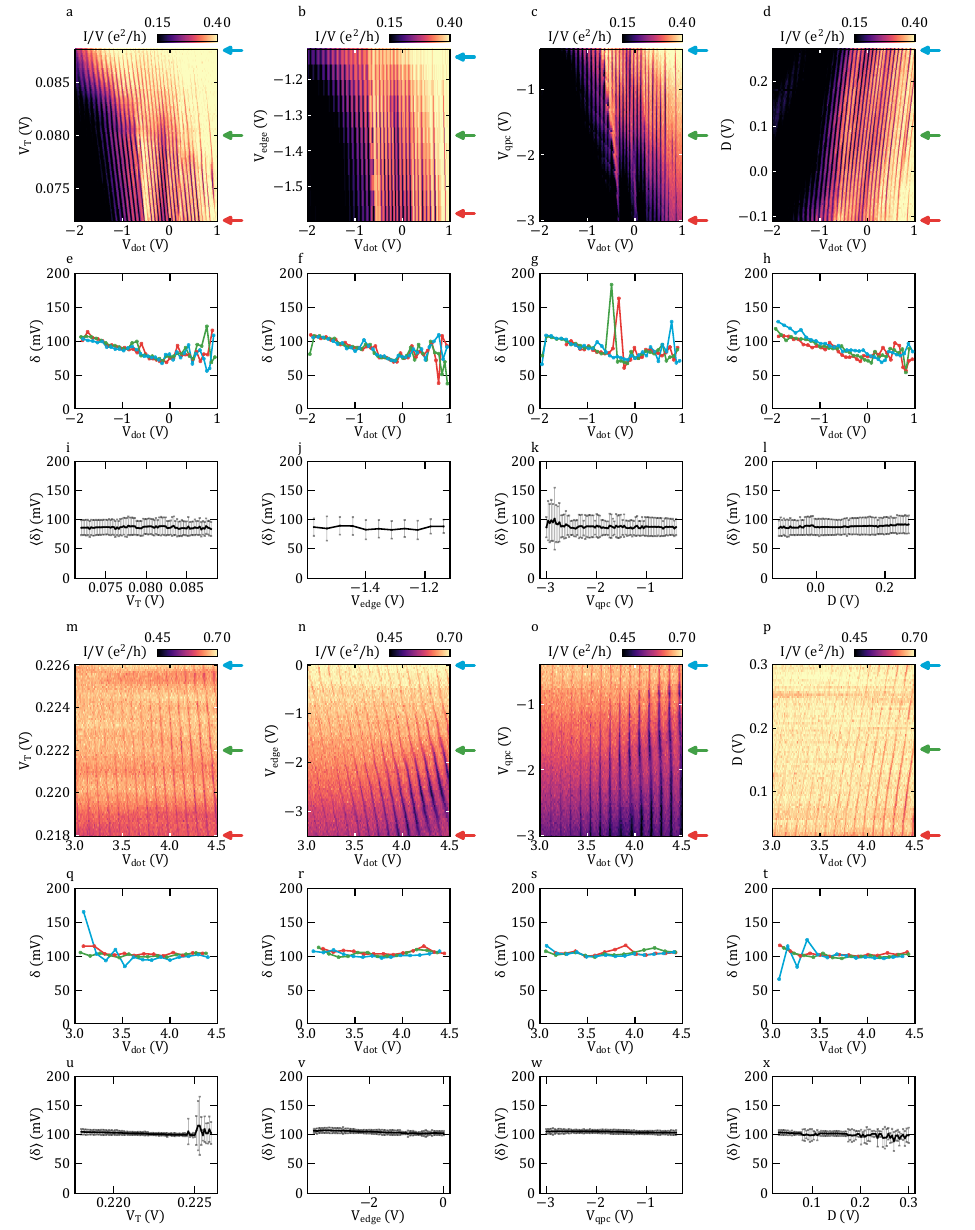} 
\caption{\textbf{Gate dependence of conductance dip spacing for $\nu = 1/3$ quasiholes and
$\nu = 2/3$ quasiparticles}.
\textbf{a--d,} Conductance $I/V$ in $\nu=1/3$ as a function of
$V_{\textrm{dot}}$ and various gates, including $V_{\textrm{T}}$,
$V_{\textrm{\rm edge}}$, $V_{\textrm{qpc}}$, and the displacement field $D$. The measurements were performed at a finite positive measured bias of approximately $V_{\textrm{bias}}\simeq 60$--$70~\mu\mathrm{V}$ after
correcting for the calibrated voltage offset. Except when swept along the slow axis, the nominal gate configurations are
$V_{\textrm{T}}=0.080~\mathrm{V}$,
$V_{\textrm{BG}}=0.30~\mathrm{V}$,
$V_{\textrm{qpc}}=-0.40~\mathrm{V}$, and
$V_{\textrm{\rm edge}}=-1.25~\mathrm{V}$. The displacement field is defined as $D=V_{\textrm{T}}-\alpha V_{\textrm{BG}}$, where $\alpha=0.628$ is the experimentally determined differential lever arm between the top and bottom gates. In panel $d$, $V_\textrm{T}$ and $V_{\textrm{BG}}$ are swept so that the electron density $n=V_{\textrm{T}}+\alpha V_{\textrm{BG}}$, where $\alpha=0.628$ is fixed.
\textbf{e--h,} Adjacent conductance dip spacings extracted from the three representative line cuts indicated in \textbf{a--d}. The blue, green, and red lines correspond to the top, middle, and bottom line cuts, respectively,
as indicated by the arrows in \textbf{a--d}.
\textbf{i--l,} Mean adjacent conductance dip spacing $\delta$ as a function of the corresponding slow-axis variable in \textbf{a--d}: $V_{\mathrm{T}}$, $V_{\mathrm{\rm edge}}$, $V_{\mathrm{qpc}}$, and $D$. At each fixed slow-axis variable value, the mean is calculated from all adjacent dip spacings along the corresponding $V_{\mathrm{dot}}$ line cut. The error bars represent one sample standard deviation of the measured spacings.
\textbf{m--x,} Similar plots are shown for $\nu=2/3$ quasiparticles. The nominal gate voltages in this configuration are:
$V_{\mathrm T}=0.220\,\mathrm V$,
$V_{\mathrm{BG}}=0.30\,\mathrm V$,
$V_{\mathrm{qpc}}=-0.40\,\mathrm V$, and
$V_{\mathrm{\rm edge}}=-1.25\,\mathrm V$.
Comparing the dip spacings across various gate configurations, we find that they are independent of the gate voltages.}
\label{fig:Spacing_Regular} 
\end{figure*}

\begin{figure*}
\centering
\includegraphics[width=5.7in]{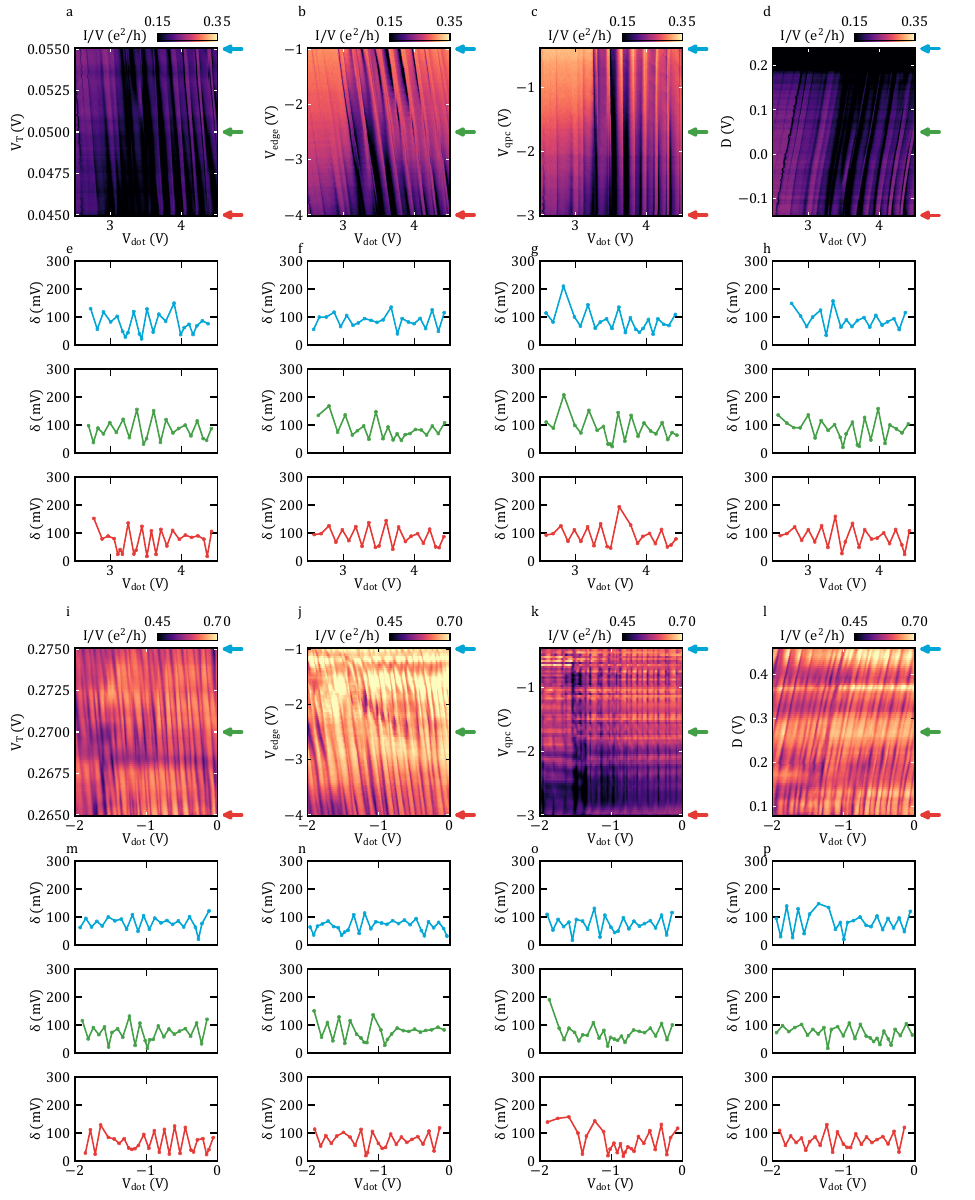} 
\caption{\textbf{Gate dependence of dip spacings for $\nu = 1/3$ quasiparticles and $\nu = 2/3$ quasiholes.}
\textbf{a-d,} Conductance $I/V$ in the $\nu=1/3$ quasiparticle regime as a function of $V_{\textrm{dot}}$ and, respectively, $V_{\textrm{T}}$, $V_{\textrm{\rm edge}}$, $V_{\textrm{qpc}}$, or the displacement field $D$. The measurements were performed at a finite positive bias, typically $V_{\textrm{bias}}\simeq 60$--$70\,\mu\mathrm{V}$. Unless being swept along the slow axis, the nominal gate voltages are
$V_{\mathrm T}=0.05\,\mathrm V$, $V_{\mathrm{BG}}=0.30\,\mathrm V$,
$V_{\mathrm{qpc}}=-3.0\,\mathrm V$, and
$V_{\mathrm{\rm edge}}=-4.0\,\mathrm V$.
The displacement field is defined as
$D=V_{\mathrm T}-\alpha V_{\mathrm{BG}}$, where $\alpha=0.628$
is the experimentally determined differential lever arm between the
top and bottom gates. In panel d, $V_{\mathrm T}$ and
$V_{\mathrm{BG}}$ are swept together while the density-control
combination $V_{\mathrm T}+\alpha V_{\mathrm{BG}}$ is held fixed.
\textbf{e--h,} Adjacent conductance dip spacings extracted from the three representative line cuts indicated in \textbf{a--d}.
\textbf{i--p,} Similar to \textbf{a-d}, but for $\nu=2/3$ quasiholes and nominal voltages are set at $V_{\mathrm T}=0.270\,\mathrm V,
V_{\mathrm{BG}}=0.30\,\mathrm V,
V_{\mathrm{qpc}}=-0.40\,\mathrm V,
V_{\mathrm{\rm edge}}=-4.0\,\mathrm V$, while the applied bias is around 60 $\mu$V. We observe an even–odd effect in all panels, although its exact locations depend strongly on the gate voltages. Irregularities in the dips also reduce the reliability of the dip-finding algorithm, which may miss resonances as they emerge.}
\label{fig:Spacing_Molecule} 
\end{figure*}

\begin{figure*}
\centering
\includegraphics[]{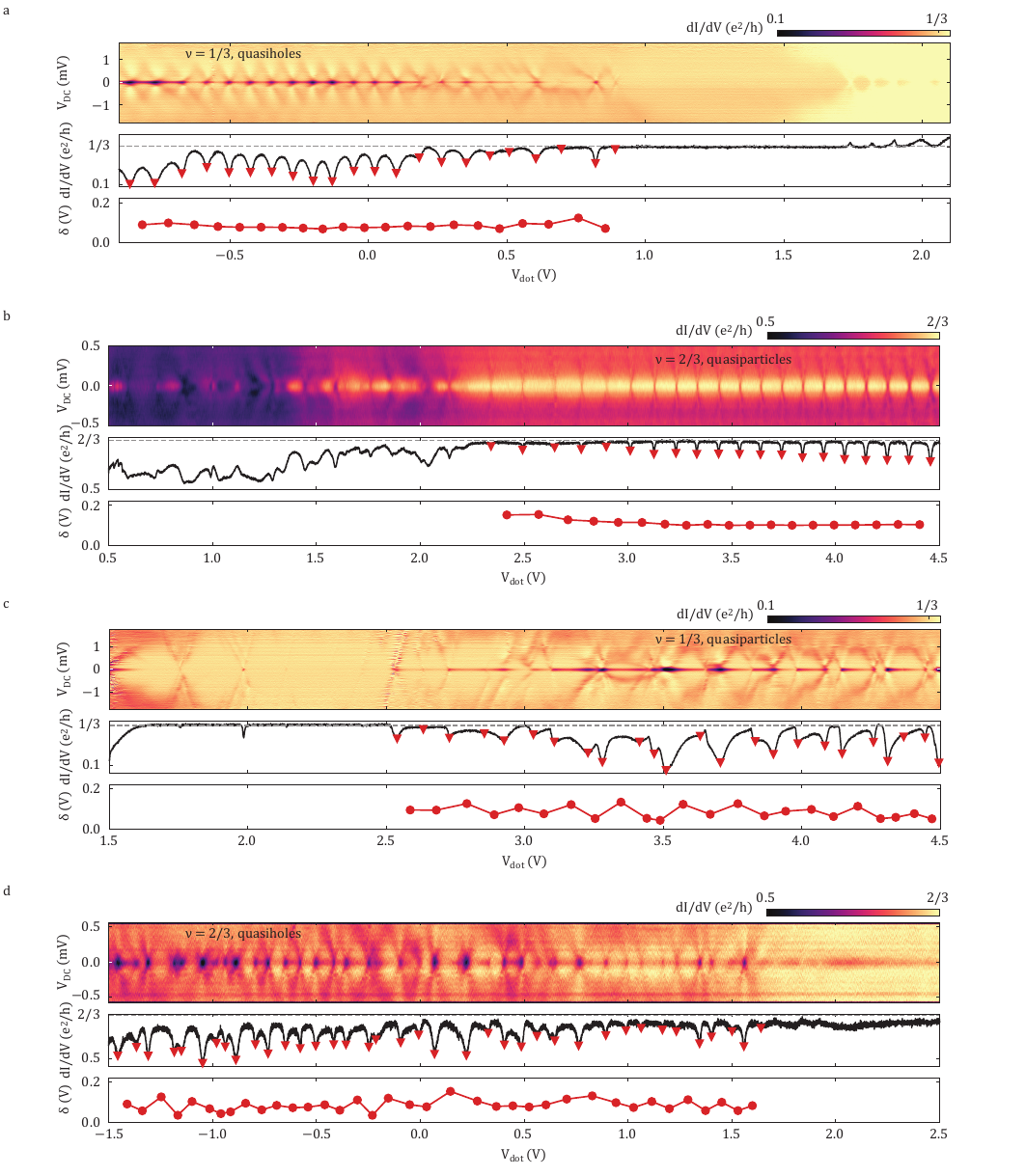} 
\caption{\textbf{Quasiparticle and quasihole Coulomb spectroscopy over an extended gate voltage range.} 
\textbf{a--d,} Extended $V_{\textrm{dot}}$-range versions of main text \ref{fig:molecular}a-d. 
In each panel, the upper plot shows the differential conductance as a function of applied bias and $V_{\textrm{dot}}$. The middle plot shows a line cut taken at zero bias. The lower plot shows $\delta$ for the dips marked by red arrows in the middle plot.}
\label{fig:Molecular_wideRange} 
\end{figure*}

\end{document}